\documentclass[english,11pt,a4paper]{article}
\usepackage{authblk}
\usepackage[text={17.0cm,23.7cm}]{geometry}
\usepackage[T1]{fontenc}
\usepackage[utf8]{inputenc}
\usepackage{babel}
\usepackage{graphicx}
\usepackage{tabularx}
\usepackage{amsmath}
\usepackage{multirow}
\usepackage{booktabs}
\usepackage{amssymb}
\usepackage[dvipsnames]{xcolor}
\usepackage{yfonts}
\usepackage{ulem}
\usepackage{enumitem}
\usepackage{comment}
\usepackage{mdframed}
\usepackage{float}
\usepackage{mathrsfs}
\usepackage{slashed}
\usepackage{tikz-feynman}
\usepackage{tikz}
\usepackage{feynmp}
\usepackage[colorlinks=true, allcolors=blue]{hyperref}

\usepackage[style=numeric-comp, sorting=none]{biblatex}
\begin{document}

	\title{\vspace{-2cm}
    {\normalsize \flushright FERMILAB-PUB-25-0955-T\\}
		\vspace{0.6cm}

	\textbf{Information-theoretic astrophysical uncertainties in the effective theory of dark matter direct detection}\\[8mm]}
	
	\author[1,2,3]{Gonzalo Herrera\thanks{gonzaloh@mit.edu}}
    \affil[1]{\normalsize\textit{Department of Physics and Kavli Institute for Astrophysics and Space Research, Massachusetts Institute of Technology, Cambridge, MA 02139, USA}}
    \affil[2]{\normalsize\textit{Harvard University, Department of Physics and Laboratory for Particle Physics and Cosmology, Cambridge, MA 02138, USA}}
    \affil[3]{\normalsize\textit{Center for Neutrino Physics, Department of Physics, Virginia Tech, Blacksburg, VA 24061, USA}}
	\date{}
	
	\maketitle
	
	\begin{abstract}
    The impact of astrophysical uncertainties in direct detection searches can vary significantly across particle dark matter models and detector targets, due to the different velocity and momentum dependencies of the scattering cross section. We address these uncertainties for all operators of the non-relativistic effective field theory of dark matter-nucleon interactions, making use of the Kullback-Leibler (KL) information divergence to measure the deviation of the true dark matter velocity distribution from the Maxwell-Boltzmann form. This approach quantifies how astrophysical uncertainties affect each operator in the effective theory, without assuming any specific functional form for the velocity distribution. While for some operators the uncertainties are smaller than one order of magnitude for entropically-motivated deviations from the Maxwell-Boltzmann form, for other operators these uncertainties can be as large as three orders of magnitude near threshold. Furthermore, we identify the dependence of the scattering rate for various operators of the effective theory with different velocity-weighted moments of the velocity distribution, functionally analogous to the mean, variance, or skewness. This provides new analytic insight into which features of the velocity distribution are most relevant to detect a given particle dark matter model. Our technique is general and could be applied to a broader class of physics problems where a physical observable depends on the statistical moments of an uncertain theoretical distribution.
	\end{abstract}

\section{Introduction}

The dark matter of the Universe may be composed by one or more fundamental particles that couple weakly to the Standard Model \cite{Pagels:1981ke,Ellis:1983ew,Jungman:1995df, Bertone:2004pz,Arcadi:2017kky,Cirelli:2024ssz}. A worldwide ongoing theoretical and experimental effort is focused on searching for dark matter particles from the Galactic halo via their scatterings off nuclei and electrons at Earth-based detectors \cite{Goodman:1984dc,Drukier:1986tm,Essig:2011nj,Schumann:2019eaa,Billard:2021uyg, Essig:2022dfa,Krnjaic:2025noj, Cheek:2025nul, Alonso-Gonzalez:2025xqg, Borah:2025wcc, SENSEI:2020dpa, DAMIC-M:2025luv,Cogswell:2021qlq, Araujo:2025rhr,QROCODILE:2024nqm,COSINUS:2017qkw, Angloher:2025yjk}. The scattering rate at the detector crucially depends on the dark matter microphysics, thus, the experimental results can be assessed systematically within an effective field theory framework for dark matter interactions with nuclei \cite{Fan:2010gt, Fitzpatrick:2012ix,Cirelli:2013ufw,Catena:2015uua,Gondolo:2020wge, Brenner:2022qku, AvisKozar:2023iyb,Cheek:2023zhv,Heimsoth:2023jgl,Drees:2024ttg, Cerdeno:2024uqt,LZ:2023lvz,DEAP:2020iwi,DarkSide-50:2020swd,CRESST:2018vwt,XENON:2022avm,PICO:2022ohk,SuperCDMS:2015lcz,Liu:2017kmx,Baumgart:2022vwr, Giffin:2025hdx}, and electrons \cite{Catena:2019gfa,He:2020wjs,Catena:2021qsr,Li:2022acp,Figueroa:2024tmn,Krnjaic:2024bdd, Giffin:2025hdx}. Another important source of uncertainty relies on the dark matter flux reaching the detector, which from simulations and data driven inference methods is not accurately known. In fact, significant deviations from the expected results by standard cosmological simulations have been suggested from kinematical tracers in the Galaxy \cite{Evans:2018bqy}, galactic stellar streams \cite{OHare:2018trr}, the large magellanic cloud \cite{Smith-Orlik:2023kyl,Reynoso-Cordova:2024xqz,Bozorgnia:2025lsl},  or non-galactic components \cite{Baushev:2012dm,Kakharov:2025myy,Herrera:2021puj, Herrera:2023fpq, Herrera:2021axy}. Capturing precisely the impact of such uncertainties is critical in the incoming era of dark matter direct detection searches that will dive into the neutrino "floor" or "fog", where deviations from the expected coherent elastic neutrino nucleus scattering cross section or solar fluxes may be degenerate with dark matter-induced signals, see \textit{e.g} \cite{Billard:2013qya,Cerdeno:2016sfi,Dutta:2017nht,AristizabalSierra:2017joc,Boehm:2018sux,Bell:2019egg,Herrera:2023xun,Carew:2023qrj,Maity:2024hzb,Demirci:2024vzk,Blanco-Mas:2024ale,Raj:2024guv,Tang:2024prl,DeRomeri:2024iaw,Dent:2025drd, Herrera:2025vjc}.

The impact of astrophysical uncertainties in the scattering rate at the detector can vary for different operators of the non-relativistic effective theory, since the dark matter-nucleon interaction cross section presents different velocity and momentum transfer dependencies for different operators. The impact of astrophysical uncertainties has been widely studied for standard spin-independent dark matter scattering \cite{Fox:2010bz, Fox:2010bu, Gondolo:2012rs, Herrero-Garcia:2012arz, DelNobile:2013cta, Frandsen:2013cna, Bozorgnia:2014gsa, Feldstein:2014gza,Blennow:2015oea, Anderson:2015xaa, Gelmini:2016pei, Kahlhoefer:2016eds, Catena:2018ywo, fowlie, Chen:2021qao,Herrera:2021axy,Herrera:2023fpq,Chen:2022xzi, Kang:2022zqv, Kang:2023gef,Bernreuther:2023aqe, Lillard:2023cyy, Lillard:2023nyj,Maity:2022enp, Folsom:2025lly, Lilie:2025wkr}, and some works have also addressed these uncertainties for other operators of the effective theory \cite{DelNobile:2013cva,Catena:2018ywo, Choi:2024rmq}, following the so-called halo-independent method \cite{Fox:2010bz,Gondolo:2012rs,DelNobile:2013cta}.

A powerful alternative to the halo-independent method was proposed in \cite{Herrera:2024zrk}. There, the deviations from the Maxwell-Boltzmann distributions were quantified via information divergences, which are commonly used in mathematics and computer science to compute the dissimilarity among two probability distributions by entropic means. The impact of such deviations on the ensuing upper limits from experiments were obtained after optimizing the recoil rate based on a  constraint on the maximal deviation from the Maxwell-Boltzmann distribution, which was set to values motivated by velocity distributions obtained from both simulations and observations. The new method from \cite{Herrera:2024zrk} was applied for several information divergences, and it was concluded that the Kullback-Leibler (KL) divergence was the most suited information divergence for this task. The analysis was performed for spin independent dark matter-nucleon and dark matter-electron interactions, and here we extend the analysis to all the operators of the effective theory of dark matter-nucleon interactions up to spin 1/2, assessing the relative impact of astrophysical uncertainties for all operators of the effective theory. The paper is organized as follows: in section \ref{sec:NREFT}, we revisit the non-relativistic effective field theory of dark matter nucleon interactions in direct detection searches. In section \ref{sec:methodology}, we present the methodology pursued, apply it to XENONnT and PICO 60 direct detection experiments \cite{XENON:2023cxc, PICO:2019vsc}, and interpret the results. Finally, in section \ref{sec:conclusions}, we present our conclusions.

\section{Direct detection of dark matter in the non-relativistic effective theory}\label{sec:NREFT}
Dark matter particles ($\chi$) from the Galactic halo would induce a differential scattering rate via nuclear ($i$) recoils at the detector, with expression
\begin{equation} \label{eq:diff_rate}
\frac{d R}{d E_R}= N_T \frac{\rho_\chi}{m_\chi} \int_{v_{\min}(E_R)} d^3 v \, f(\vec{\bf{v}})\, v \frac{d \sigma_{\chi-N}}{d E_R},
\end{equation} Here $N_T$ denotes the number of target nuclei per unit detector mass, $v \equiv\left|\vec{\mathbf{v}}\right|$ is the dark matter velocity in the laboratory frame, $v_{\min }=\sqrt{\frac{m_N E_R}{2 \mu_{\chi-N}^2}}$ is the minimum dark matter velocity necessary to induce producing a recoil with energy $E_R$, $\mu_{\chi-N}$ the dark matter-nucleus reduced mass and $m_N$ the nucleus mass. The astrophysics part comes from $\rho_\chi$ and $f\left(\vec{\mathbf{v}}\right)$, denoting the local dark matter density and velocity distribution respectively. For the local density we take $\rho_\chi=0.3 \, \mathrm{GeV} / \mathrm{cm}^3$ \cite{Read:2014qva}. The velocity distribution will be the source of uncertainty addressed in this work. For the "benchmark" distribution, over which deviations will be measured, we consider a Maxwell-Boltzmann distribution
\begin{align}\label{eq:MB}
    		f_\text{MB}(\vec{\mathbf{v}}) = \frac{1}{N\,(2\pi\,\sigma_v^2)^{3/2}}\,\exp\left(-\frac{\vec{\mathbf{v}}^2}{2\,\sigma_v^2}\right)\,,
\end{align}
with velocity dispersion $\sigma_v\approx 156\,\text{km/s}$ \cite{Kerr:1986hz, Green:2011bv}.
The normalization constant $N$ depends on the escape velocity $v_\text{esc}\approx 544\,\text{km/s}$ \cite{Smith:2006ym, Piffl:2013mla, Necib:2021vxr, Necib:2021yhq} and is given by
\begin{align}
    		N = \text{erf}\left(\frac{v_\text{esc}}{\sqrt{2}\sigma_v}\right)-\sqrt{\frac{2}{\pi}}\,\frac{v_\text{esc}}{\sigma_v}\,\exp\left(-\frac{v_\text{esc}^2}{2\sigma_v^2}\right)\,.
\end{align}
In the following, we consider the angular-averaged 3D distributions in the solar rest frame
\begin{align}
    f(v) = \int\text{d}\Omega_v\, f(\vec{\mathbf{v}}+\vec{\mathbf{v}}_\odot)\,,
    \label{eq:angular_average}
\end{align}
with $\vec{\mathbf{v}}_\odot$ the velocity of the Sun with respect to the Galactic rest frame. We emphasise that the symbol $f$ is used with two distinct meanings throughout, distinguished only by the argument: the three-dimensional velocity distribution $f(\mathbf v)$ in Eq.~(\ref{eq:diff_rate}), and the angular-averaged speed distribution $f(v)$ defined by Eq.~(\ref{eq:angular_average}), which is the object entering the velocity-weighted moment integrals and the KL divergence. The velocity of the Sun depends on the motion of the local standard of rest (LSR) and the peculiar motion of the Sun with respect to the LSR, i.e.
\begin{align}
    		\vec{\mathbf{v}}_\odot = \vec{\mathbf{v}}_\text{LSR} + \vec{\mathbf{v}}_{\odot, \text{pec}}\,.
\end{align}
The motion of the LSR is given by $\vec{\mathbf{v}}_\text{LSR}=(0,v_c,0)$ where $v_c\approx 220\,\text{km/s}$ \cite{Kerr:1986hz} is the local circular speed. Moreover, the Sun's peculiar motion is $\vec{\mathbf{v}}_{\odot, \text{pec}} = (11.1, 12.24, 7.25)\,\text{km/s}$ \cite{Schoenrich2010}. Here we will consider all isotopes present at the XENONnT \cite{XENON:2023cxc} and the PICO60 \cite{PICO:2019vsc} experiments with their corresponding relative abundances.

The particle physics-dependent part in Eq. \ref{eq:diff_rate} is encoded in the differential scattering cross section $\frac{d \sigma_{\chi-N}}{d E_R}$. In an effective theory description, the interacting Hamiltonian between the dark matter $\chi$ with spin up to 1/2, and the nucleus $N$, reads \cite{haxton1, haxton2}

\begin{equation}
\mathcal{H}(\mathbf{r})=\sum_{\tau=0,1} \sum_{j=1}^{15} c_j^\tau O_j(\mathbf{r}) t^\tau.
\end{equation}
Here, $O_j$ denotes the operators of the effective theory, and $c_j$ the corresponding Wilson coefficients. $t^0$ and $t^1=\tau_3$ denote the $2\times 2$ identity and the third Pauli matrix in isospin space, respectively. A list of all the Galilean invariant operators is provided in Table~\ref{tab:operators}. These depend on $1_{\chi N}$, which is the identity operator, the transferred momentum $\vec{q}$, $\vec{S}_X$ and $\vec{S}_N$ are the dark matter and nucleon spins, respectively, while $\vec{v}^{\perp}=\vec{v}+\frac{\vec{q}}{2 \mu_{\chi N}}$ is the relative transverse velocity operator satisfying $\vec{v}^{\perp} \cdot \vec{q}=0$.

\begin{table}[t!]
\begin{center}
\begin{tabular}{|l|l|}
\hline
$ \mathcal{O}_1 = 1_\chi 1_N$ & $\mathcal{O}_9 = i \vec{S}_\chi \cdot (\vec{S}_N \times \frac{\vec{q}}{m_N})$ \\
$\mathcal{O}_3 = i \vec{S}_N \cdot (\frac{\vec{q}}{m_N} \times \vec{v}^\perp)$ & $\mathcal{O}_{10} = i \vec{S}_N \cdot \frac{\vec{q}}{m_N}$ \\
$\mathcal{O}_4 = \vec{S}_\chi \cdot \vec{S}_N$ & $\mathcal{O}_{11} = i \vec{S}_\chi \cdot \frac{\vec{q}}{m_N}$\\
$\mathcal{O}_5 = i \vec{S}_\chi \cdot (\frac{\vec{q}}{m_N} \times \vec{v}^\perp)$ & $\mathcal{O}_{12} = \vec{S}_\chi \cdot (\vec{S}_N \times \vec{v}^\perp)$ \\
$\mathcal{O}_6= (\vec{S}_\chi \cdot \frac{\vec{q}}{m_N}) (\vec{S}_N \cdot \frac{\vec{q}}{m_N})$ & $\mathcal{O}_{13} =i (\vec{S}_\chi \cdot \vec{v}^\perp  ) (  \vec{S}_N \cdot \frac{\vec{q}}{m_N})$ \\
$\mathcal{O}_7 = \vec{S}_N \cdot \vec{v}^\perp$ & $\mathcal{O}_{14} = i ( \vec{S}_\chi \cdot \frac{\vec{q}}{m_N})(  \vec{S}_N \cdot \vec{v}^\perp )$\\
$\mathcal{O}_8 = \vec{S}_\chi \cdot \vec{v}^\perp$ & $\mathcal{O}_{15} = - ( \vec{S}_\chi \cdot \frac{\vec{q}}{m_N}) \big((\vec{S}_N \times \vec{v}^\perp) \cdot \frac{\vec{q}}{m_N}\big)$ 
\\ \hline
\end{tabular}
\caption{Non-relativistic Galilean invariant operators for dark matter-nucleus interactions for dark matter spin up to $1/2$.}
\label{tab:operators}
\end{center}
\end{table}

The differential scattering cross section induced by each operator is given by
\begin{equation}
\frac{d \sigma_{\chi-N}}{d E_R}=\frac{2 m_N}{4 \pi v^2}\left[\frac{1}{2 j_{\chi}+1} \frac{1}{2 j_N+1}\left|\mathcal{M}_{\chi-N}\right|^2\right]
\end{equation}
where $j_{\chi}, j_{N}$ are the dark matter and nucleus N spins. The matrix element for the scattering process can be expressed as a linear combination of dark matter and nuclear "response functions", for which we follow the formalism and notation of \cite{haxton1, haxton2}
\begin{equation}
\frac{1}{2 j_{\chi}+1} \frac{1}{2 j_N+1}\left|\mathcal{M}_{\chi-N}\right|^2=\frac{4 \pi}{2 j_N+1} \sum_{\tau=0,1} \sum_{\tau^{\prime}=0,1} \sum_k R_k^{\tau \tau^{\prime}}\left[c_j^\tau,\left(v^{\perp}\right)^2, \frac{q^2}{m_N^2}\right] W_{N k}^{\pi \tau^{\prime}}(y).
\end{equation}
Here, $c_j$ are the Wilson coefficients for each operator $j$. $W_{N k}^{\tau \tau^{\prime}}(y)$ denotes the nuclear response functions for each operator $k$, and $y=q^2b^2/4$, with $b$ the size of the nucleus. $R_k^{\tau \tau^{\prime}}$ denotes the dark matter response functions. There is a correspondence between each dark matter operator $j$ and the corresponding nuclear operators $k$. Such correspondence can be better understood by decomposing the nuclear response function via \cite{Kang:2018odb}
\begin{equation}
R_k^{\tau \tau^{\prime}}=R_{0 k}^{\tau \tau^{\prime}}+R_{1 k}^{\tau \tau^{\prime}}\left(v^{\perp}\right)^2=R_{0 k}^{\tau \tau^{\prime}}+R_{1 k}^{\tau \tau^{\prime}}\left(v^2-v_{\min }^2\right)\textcolor{magenta}{,}
\end{equation}
Under this decomposition, it is possible to map which nuclear operators affect each Wilson coefficient, see Table~\ref{tab:nuclear_operators}. The six nuclear response functions appearing in Table~\ref{tab:nuclear_operators} were defined in Refs.~\cite{haxton1,haxton2,Fitzpatrick:2012ix}. $M(q^2)$ is the standard spin-independent (coherent charge) response; $\Sigma'(q^2)$ and $\Sigma''(q^2)$ are the transverse and longitudinal components of the spin response and dominate the usual spin-dependent scattering; $\Phi''(q^2)$ and $\widetilde\Phi'(q^2)$ couple the nucleon spin to its orbital angular momentum, and appear for operators involving the momentum transfer $\vec q$ or the transverse velocity $\vec v^{\perp}$; and $\Delta(q^2)$ is the convection (orbital) response, generated by operators involving $\vec v^{\perp}$ alone. The $q$- and $v$-dependent prefactors accompanying each response function in Table~\ref{tab:nuclear_operators} are a direct consequence of this structure and reflect how strongly the corresponding operator probes the high-velocity tail of the dark matter distribution.

\begin{table}[t!]
\centering

\begin{tabular}{|c|c|c|}
\hline
Operator & $R_{0 k}^{\tau \tau}$ & $R_{1 k}^{\tau \tau'}$ \\
\hline
$\mathcal{O}_1$   & $M(q)$  & -- \\
$\mathcal{O}_3$   & $\Phi''(q^4)$ & $\Sigma'(q^2)$ \\
$\mathcal{O}_4$   & $\Sigma''(q),\, \Sigma'(q)$ & -- \\
$\mathcal{O}_5$   & $\Delta(q^4)$ & $M(q^2)$ \\
$\mathcal{O}_6$   & $\Sigma''(q^4)$ & -- \\
$\mathcal{O}_7$   & -- & $\Sigma'(q)$ \\
$\mathcal{O}_8$   & $\Delta(q^2)$ & $M(q)$ \\
$\mathcal{O}_9$   & $\Sigma'(q^2)$ & -- \\
$\mathcal{O}_{10}$ & $\Sigma''(q^2)$ & -- \\
$\mathcal{O}_{11}$ & $M(q^2)$ & -- \\
$\mathcal{O}_{12}$ & $\Phi''(q^2),\, \widetilde{\Phi}'(q^2)$ 
                    & $\Sigma''(q),\, \Sigma'(q)$ \\
$\mathcal{O}_{13}$ & $\widetilde{\Phi}'(q^4)$ & $\Sigma''(q^2)$ \\
$\mathcal{O}_{14}$ & -- & $\Sigma'(q^2)$ \\
$\mathcal{O}_{15}$ & $\Phi^{\prime \prime}\left(q^6\right)$ & $\Sigma'(q^4)$ \\
\hline
\end{tabular}
\caption{Nuclear response functions for the non-relativistic dark matter-nuclei interaction operators. Following the convention of Ref.~\cite{Fitzpatrick:2012ix}, the operator $\mathcal{O}_2$ is not included in the basis employed in this work, as it does not arise at leading order in the non-relativistic reduction of any renormalisable Lorentz-invariant interaction.}\label{tab:nuclear_operators}
\end{table}

\iffalse
\begin{table}[t!]
\centering
$\begin{array}{|c|c|c|} \label{tab:nuclear_operators}
\mathrm{ Operator } & R_{0 k}^{\tau \tau} & R_{1 k}^{\tau \tau^{\prime}} \\
\hline \mathcal{O}_1 & M\left(q\right) & - \\
\mathcal{O}_3 & \Phi^{\prime \prime}\left(q^4\right) & \Sigma^{\prime}\left(q^2\right) \\
\mathcal{O}_4 & \Sigma^{\prime \prime}\left(q\right), \Sigma^{\prime}\left(q\right) & - \\
\mathcal{O}_5 & \Delta\left(q^4\right) & M\left(q^2\right) \\
\mathcal{O}_6 & \Sigma^{\prime \prime}\left(q^4\right) & - \\
\mathcal{O}_7 & - & \Sigma^{\prime}\left(q\right) \\
\mathcal{O}_8 & \Delta\left(q^2\right) & M\left(q\right) \\
\mathcal{O}_9 & \Sigma^{\prime}\left(q^2\right) & - \\
\mathcal{O}_{10} & \Sigma^{\prime \prime}\left(q^2\right) & - \\
\mathcal{O}_{11} & M\left(q^2\right) & - \\
\mathcal{O}_{12} & \Phi^{\prime \prime}\left(q^2\right), \widetilde{\Phi}^{\prime}\left(q^2\right) & \Sigma^{\prime \prime}\left(q\right), \Sigma^{\prime}\left(q\right)\\
\mathcal{O}_{13} & \tilde{\Phi}^{\prime}\left(q^4\right) & \Sigma^{\prime \prime}\left(q^2\right)\\
\mathcal{O}_{14} & - & \Sigma^{\prime}\left(q^2\right)\\
\hline
\end{array}$
\end{table}
\fi

\section{Methodology and results}\label{sec:methodology}

The method derived in \cite{Herrera:2024zrk} can be used to bound the impact of astrophysical uncertainties on the non-relativistic effective theory of dark matter-nucleon interactions. Some works in the literature have been devoted to similar goals in the past \cite{Kahlhoefer:2016eds,Catena:2018ywo,Kang:2022zqv}, but with traditional halo-independent methods, mainly focused on interpreting signals from a given experiment given constraints from different ones, and without taking into account all the operators of the effective theory.

Here we derive constraints on the dark matter-nucleon scattering cross section from the XENONnT and PICO60 experiments, for all of the operators in the effective theory, and quantifying the deviation from the Maxwell-Boltzmann velocity distribution with a KL-divergence, defined as \cite{Kullback51klDivergence}
\begin{equation}
D_{\mathrm{KL}}\!\left[f_{\mathrm{MB}}\,\|\,f\right]
=
\int d^3v \;
f_{\mathrm{MB}}(\vec{\mathbf{v}})\,
\log\!\left(\frac{f_{\mathrm{MB}}(\vec{\mathbf{v}})}{f(\vec{\mathbf{v}})}\right),
\end{equation}

where both \(f_{\mathrm{MB}}(\vec{\mathbf{v}})\) and \(f(\vec{\mathbf{v}})\) are consistently normalized. The KL-divergence provides an information-theoretic measure of how distinguishable a distribution \(f\) is from a reference distribution \(f_{\rm MB}\), quantifying the amount of information lost when \(f_{\rm MB}\) is used to approximate the true velocity distribution \(f\)\footnote{The KL-divergence, besides measuring the relative entropy among two distributions, is also directly connected to common metrics of fine-tuning used in related particle physics contexts, see \cite{Fowlie:2024nhs}.}. Here \(D_{\rm KL}\) offers a model-independent and basis-invariant metric to assess departures from the standard Maxwell-Boltzmann halo assumption and to evaluate their impact on direct detection constraints.

We choose maximally-allowed values of the KL-divergence motivated by calculations of the dark matter velocity distribution (galactically-bounded only) from simulations and observations \cite{Vogelsberger:2008qb, Evans:2018bqy, Necib:2018iwb, Schaye:2014tpa,OHare:2018trr}. The value $D_{\rm KL}=0.1$ adopted in this work represents a conservative benchmark for the allowed deviation from the Maxwell--Boltzmann reference. For comparison, velocity distributions extracted from cosmological simulations (Aquarius, Eagle) and from Gaia-motivated models (SHM$^{++}$, S1-stream+SHM) typically differ from the Maxwell--Boltzmann form at the level $D_{\rm KL}\sim 0.01-0.1$ \cite{Herrera:2024zrk}, so $D_{\rm KL}=0.1$ encompasses the bulk of entropically plausible halo models. We also explore $D_{\rm KL}=1$ as a more aggressive benchmark to bracket extreme astrophysical scenarios. We then solve the following optimization problem for the experiment $\mathscr E$, with free variables the Wilson coefficients of the effective theory $\vec{c}$ and velocity distribution $f(\vec{\mathbf{v}})$:

\begin{center}
\begin{align}\label{eq:opt_problem_NREFT_VDF}
& N_{\rm opt }^{\mathscr E} \equiv \min / \max _{f(\vec{\mathbf{v}})}\left[N_{f(\vec{\mathbf{v}})}^{\mathscr E}(\mathbf{c})\right], \\
& \text { subject to } D_{\rm KL}[f_{\rm MB}(\vec{\mathbf{v}})\,\|\,f(\vec{\mathbf{v}})] \leq K, \\
& \text { and } \int f(\vec{\mathbf{v}}) \mathrm{d}^3 v=1,
\end{align}
\end{center}
using convex optimization techniques, and the code \texttt{CVXPY} \cite{diamond2016cvxpy}, interfaced with the \texttt{ECOS} conic solver~\cite{domahidi2013ecos}. After discretising $f(v)$ on a velocity grid, the normalisation and non-negativity conditions become standard linear constraints on the grid values, while the KL-divergence bound $D_{\rm KL}(f_{\rm MB}\|f)\leq K$ is encoded through a set of exponential-cone constraints, see Appendix A in Ref. \cite{Herrera:2024zrk}. \texttt{ECOS} then solves the resulting conic programme by a primal--dual interior-point method, which iteratively approaches the optimum from inside the feasible region by solving a linearised form of the optimality conditions at each step. The distribution $f_\star(v)$ that saturates this optimisation at a given $K$ is not universal, rather it depends both on the dark matter mass, through the kinematic threshold $v_{\min}(m_\chi)$, and on the specific effective-theory operator, through the $v$- and $q$-weighting of the integrand. A different $f_\star(v)$ is therefore obtained for each (operator,\,$m_\chi$) pair, and the shape of $f_\star(v)$ is itself a diagnostic of which features of the velocity distribution a given operator is most sensitive to. Velocity-independent operators are driven by the bulk of the distribution, while operators with stronger velocity- or momentum-weightings concentrate $f_\star(v)$ in the high-velocity tail. The optimized $f_\star(v)$ for a range of dark matter masses can vary significantly, see Ref. ~\cite{Herrera:2024zrk}.

We solve the problem for the experiments XENONnT and PICO60, employing an adapted version of the code \texttt{WimPyDD} to compute the scattering rates at these experiments \cite{Jeong:2021bpl, Kang:2025yci}. The choice of this ordering, rather than the
reverse \(D_{\mathrm{KL}}[f\,\|\,f_{\mathrm{MB}}]\), is not arbitrary. It is what makes the variational problem that defines our exclusion bands (minimising and maximising the event-rate functional
\(N^{\mathscr E}[f]\) subject to a bound on the KL divergence, at fixed
\(f_{\mathrm{MB}}\)) convex in the unknown speed distribution \(f(v)\) \cite{Herrera:2024zrk}. Here, $K$ denotes the maximal value of the KL-divergence between the reference (Maxwell-Boltzmann) and optimized distribution. We consider in the following the values $K=0.1, 1$.

It becomes important to point out that the effective theory contains several operators depending on different nuclear response functions. These depend differently on the momentum transfer of the scattering process, which enters crucially in the calculation of the experimental rates, and affects the impact of the velocity distribution on these. In particular, the dark matter response function $R^{\tau\tau^{\prime}}_{k}$ can be decomposed as~\cite{haxton2},
\begin{equation}
R_k^{\tau\tau^{\prime}}=R_{0k}^{\tau\tau^{\prime}}+R_{1k}^{\tau\tau^{\prime}}
(v^{\perp}_T)^2=R_{0k}^{\tau\tau^{\prime}}+R_{1k}^{\tau\tau^{\prime}}\left
(v_T^2-v_{\rm min}^2\right ),
\label{eq:r_decomposition}
\end{equation}
and each operator is dominated by a different response function, with with different momentum and velocity dependencies. 

We can classify in advance the operators according to their momentum and velocity dependence in the following way:

\begin{itemize}

\item $\mathcal{O}_1$ and $\mathcal{O}_{4}$ only present a velocity independent DM response function $R_{0k}^{\tau\tau^{\prime}}$, and their nuclear response functions are momentum-independent $\propto q^{0}$. The impact of astrophysical uncertainties is therefore expected to be qualitatively similar for the canonical spin-independent and spin-dependent operators.

\item $\mathcal{O}_3$ and $\mathcal{O}_{6}$ present a velocity independent DM response function with nuclear response functions which scale as $\propto q^4$. $\mathcal{O}_3$ also presents a velocity-dependent DM response function $R_{1k}^{\tau\tau^{\prime}}$ depending on $\Sigma^{'}(q^2)$. This will only be mildly affected by the velocity distribution compared to the velocity-independent term in Xenon, but may play an important role on Fluorine. This is because the velocity-independent term of $\mathcal{O}_3$ is $\Phi^{\prime \prime}$, which is sensitive to ($\vec{L} \cdot \vec{S}$), which favours non-completely filled angular momentum orbitals, and therefore heavy elements like Xenon. For Fluorine, the spin-dependent response  $\Sigma^{'}$ might dominate due to the unpaired nucleon.

\item $\mathcal{O}_5$ and $\mathcal{O}_{14}$ present a velocity-dependent DM response function with nuclear response functions scaling as $\propto q^2$. $\mathcal{O}_5$ also presents a velocity-independent DM response function with nuclear response function $\Delta(q^4)$, which is however subdominant with respect to the velocity-independent DM response function, which depends on $M(q^2)$. The $\Delta$ response measures the nucleon angular momentum content of the nucleus, and can be important for ${}^{19}$F or the isotope $^{131}$Xe, which have an unpaired nucleon in a non s-shell orbital. However, the coherent response function $M$ couples to the total number of nucleons and it is dominant for Xenon and even Fluorine.

\item $\mathcal{O}_7$ and $\mathcal{O}_{8}$ present velocity dependent DM response functions with momentum-independent nuclear response functions ($ \propto q^0$). $\mathcal{O}_{8}$ also presents a velocity-independent DM response function with nuclear response function $\Delta(q^2)$, which is however subdominant with respect to the velocity-independent DM response function, which depends on $M(q^0)$. 

\item $\mathcal{O}_{9}$, $\mathcal{O}_{10}$, $\mathcal{O}_{11}$, $\mathcal{O}_{12}$ present velocity-independent response functions, with nuclear response functions scaling as $\propto q^2$. $\mathcal{O}_{12}$ also presents a velocity-dependent DM response function with momentum-independent nuclear response functions $\Sigma^{\prime \prime}(q^0)$ and $\Sigma^{\prime}(q^0)$. These are subdominant w.r.t. the velocity-independent ones $\Phi^{\prime \prime}(q^2)$ and $\Phi^{\prime}(q^2)$ for xenon, but they may be dominant for Fluorine

\item $\mathcal{O}_{13}$ is an independent species, with both velocity-independent and velocity-dependent DM response functions, with nuclear response functions with dependencies $\propto q^2$ and $\propto q^4$, respectively.

\item $\mathcal{O}_{15}$ is unique in the sense that it is the only operator presenting a velocity-independent DM response function with nuclear response function with $\propto q^6$ dependence, and a velocity-dependent one with $\propto q^4$.

\end{itemize}

We can interpret the impact of the velocity distribution on the scattering rate for different operators in terms of velocity-weighted integrals of the dark matter speed distribution. Here and throughout, $N^{\mathscr E}$ denotes the expected event rate (per unit target-nucleus number and per unit dark matter density) generated by a given effective-theory operator, so that the differential rate of Eq.~(\ref{eq:diff_rate}) is proportional to $N^{\mathscr E}$ once the $q$-dependent nuclear form factors are integrated out.
\begin{equation}
N^{\mathscr E} \propto \int_{v \geq v_{\rm min}} dv \, f(v)\, v^n ,
\end{equation}
where the power \(n\) is determined by the velocity and momentum dependence of the corresponding non-relativistic operator. It is immediately noticeable that these integrals appear similar to the definitions of the statistical moments of a distribution, such as the mean ($n=1$), the variance ($n=2$), the skewness ($n=3$) and the kurtosis ($n=4$). However, these integrals should be understood as truncated raw moments of the speed distribution, rather than as statistical moments in the usual sense. The functional form indeed coincides with that of the raw statistical moments, but the presence of a lower cutoff \(v_{\min}\) and the ambiguous normalization preclude a direct statistical interpretation. We refer the reader to Appendix \ref{app:moments} for a discussion on the mapping of the velocity-weighted integrals into the statistical moments of the distribution.

The integration is not performed over the full domain of velocities available to dark matter particles, but instead begins at the minimum velocity required to induce a detectable nuclear recoil,
\begin{equation}
v_{\min} = \sqrt{\frac{m_N E_R}{2 \mu_{\chi-N}^2}},
\end{equation}
which depends on the momentum transfer of the scattering. For intermediate and heavy dark matter masses, \(m_\chi \gtrsim 20~\mathrm{GeV}\), the accessible velocity range covers a large fraction of the distribution, and the rate is therefore sensitive to its bulk properties, allowing for a cleaner comparison with the statistical moments. For lighter dark matter masses, however, the rate becomes increasingly dominated by the high-velocity tail of the distribution.

The velocity dependence of the event rates for each operator in the effective theory can be made explicit through the integrated dark matter response functions,
\begin{equation}
N^{\mathscr E}_{\mathcal O_i} \propto \int_{v_{\min} = \frac{q}{2\mu_T}} dv \, v f(v)
\int_0^{\infty} dq \, q \, F(v,q),
\end{equation}
where \(\mu_T\) is the reduced mass of the dark matter--target system and \(F(v,q)\) denotes the form factors encoding the momentum and velocity dependence of the scattering, as determined by both the dark matter and nuclear response functions (see Table~\ref{tab:nuclear_operators}).

In what follows we organise the discussion by operator family, grouping the effective-theory operators according to the velocity- and momentum-weighting structure that each of them imparts on the integrand in the equation above. We consider in turn: (i) velocity-independent operators, such as $\mathcal O_1$ and $\mathcal O_4$, which probe the lowest-order velocity moments of $f(v)$ and are therefore primarily sensitive to the bulk of the speed distribution; (ii) operators with explicit $\vec v^{\perp}$ dependence in the dark matter response function, such as $\mathcal O_3$ and $\mathcal O_{12}$, which probe higher raw moments of $f(v)$; and (iii) operators with higher $q$-power suppression, such as $\mathcal O_6$, $\mathcal O_{11}$ and $\mathcal O_{15}$, whose rates are dominated by the high-velocity tail of the distribution. Within each family the sensitivity to astrophysical uncertainties scales with the effective velocity weighting of the integrand, as made quantitative in the paragraphs below.

For the standard spin-independent and spin-dependent operators, the dark matter response function contains no explicit velocity dependence beyond the kinematic prefactor, and is also momentum independent.
In this case,
\begin{equation}
N^{\mathscr E}_{\mathcal O_1} \propto \int_{v_{\min}} dv \, v f(v),
\end{equation}
so that the rate is primarily sensitive to the lower-order velocity weighting of the distribution. This weighting resembles that entering the definition of the mean speed for an untruncated distribution, and consequently the predicted event rate is most affected by modifications to the bulk of the velocity distribution relative to the Standard Halo Model.

In contrast, operators with stronger velocity dependence probe higher-order velocity weightings. For instance, operators whose dark matter response functions contain an explicit dependence on \(v_\perp^2\), such as \(\mathcal O_3\) and \(\mathcal O_{12}\), lead to velocity-weighted integrals proportional to the second raw moment of the speed distribution. While this weighting is formally analogous to the second raw moment entering the definition of the variance for an untruncated distribution, it does not correspond to the variance itself, but instead reflects sensitivity to a combination of the bulk and moderately high-velocity regions of the distribution, see Appendix \ref{app:moments}.

Another example are the operators \(\mathcal O_5\) and \(\mathcal O_8\), dominated by a velocity-dependent dark matter response function combined with a coherent nuclear response, yielding
\begin{equation}
N^{\mathscr E}_{\mathcal O_5} \propto \int_{v_{\min}} dv \, v^3 f(v).
\end{equation}
The cubic velocity weighting is formally analogous to that appearing in the raw moments that contribute to the definition of skewness in an untruncated distribution. The skewness measures the asymmetry of the probability distribution around its mean, so this operator will be mostly sensitive to changes in the tails of the velocity distribution.

\begin{figure}[H]
\begin{minipage}[H]{0.49\linewidth} % Adjust the width as desired
\includegraphics[width=\linewidth]{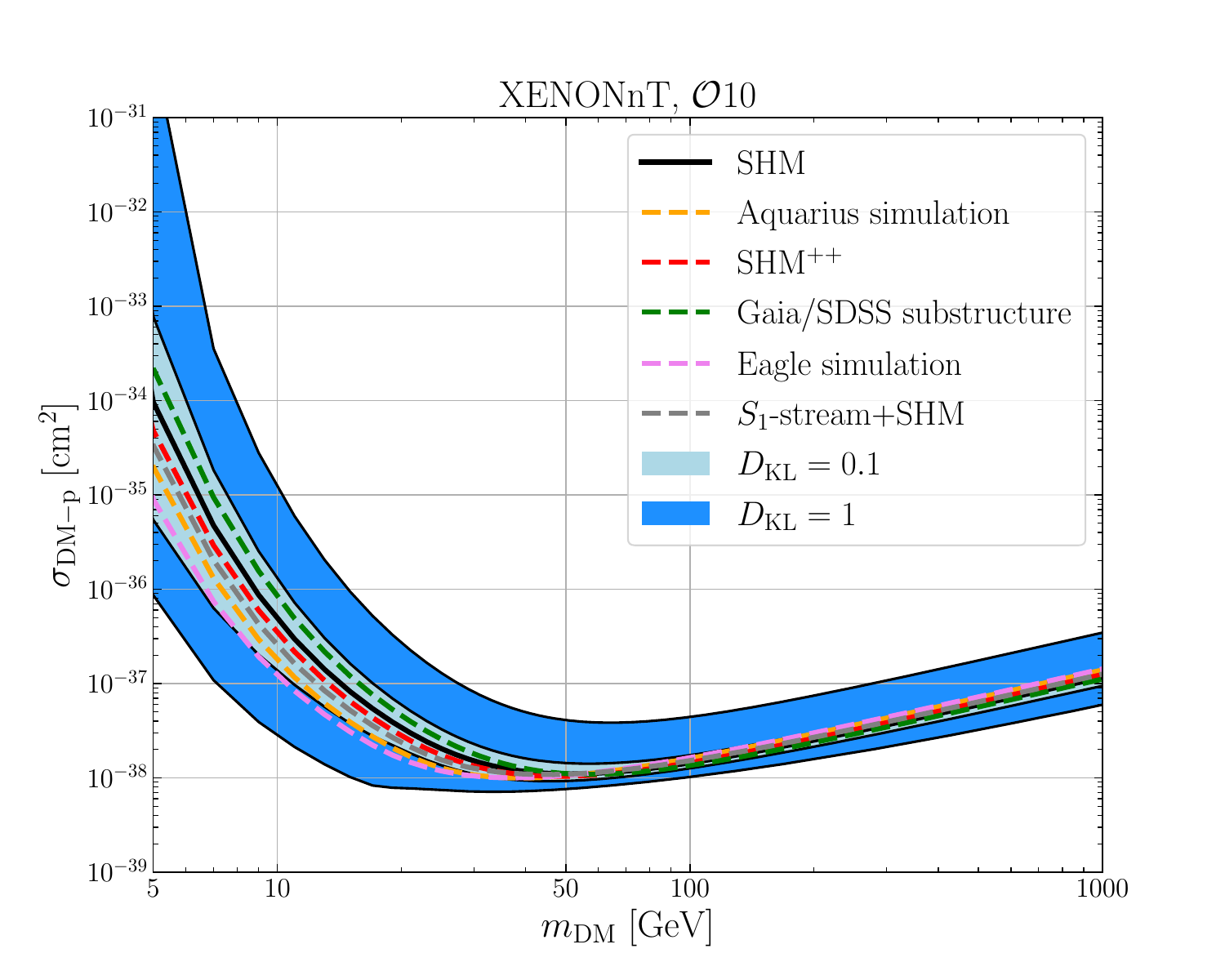}
\end{minipage}\hfill
\begin{minipage}[H]{0.49\linewidth} % Adjust the width as desired
\includegraphics[width=\linewidth]{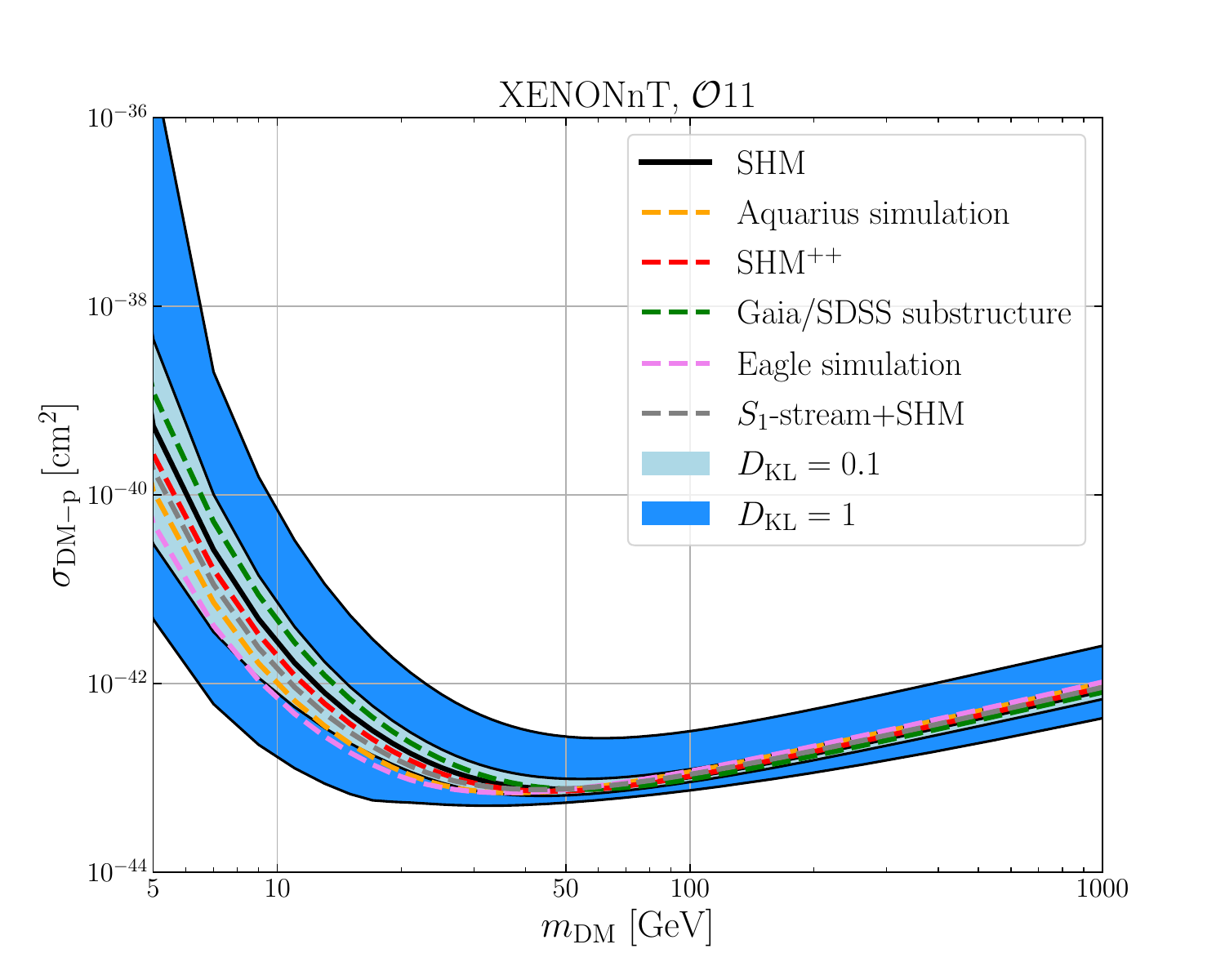}
\end{minipage}

\vspace{-2mm} % Adjust the vertical spacing as desired

\begin{minipage}[H]{0.49\linewidth} % Adjust the width as desired
\includegraphics[width=\linewidth]{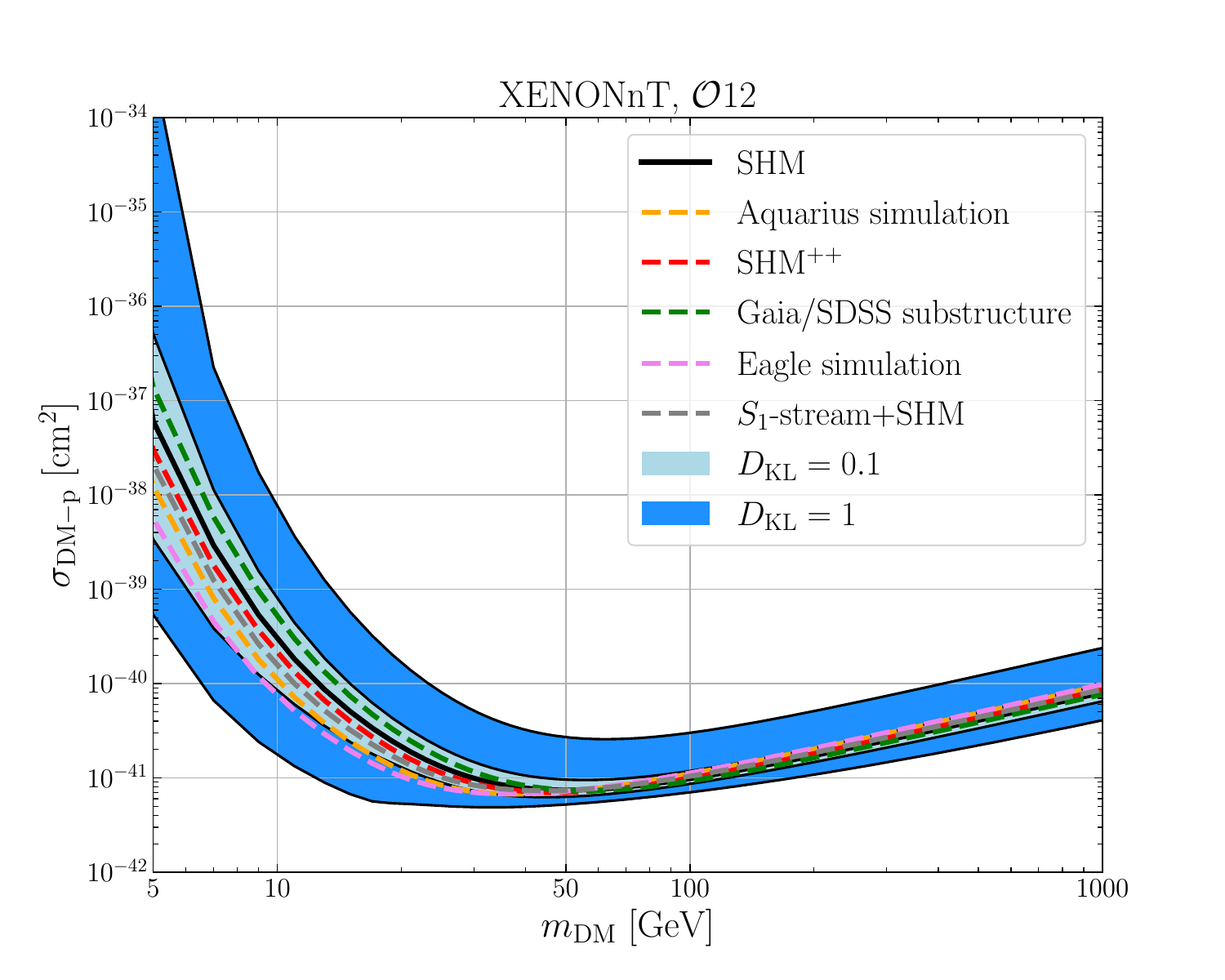}
\end{minipage}\hfill
\begin{minipage}[H]{0.49\linewidth} % Adjust the width as desired
\includegraphics[width=\linewidth]{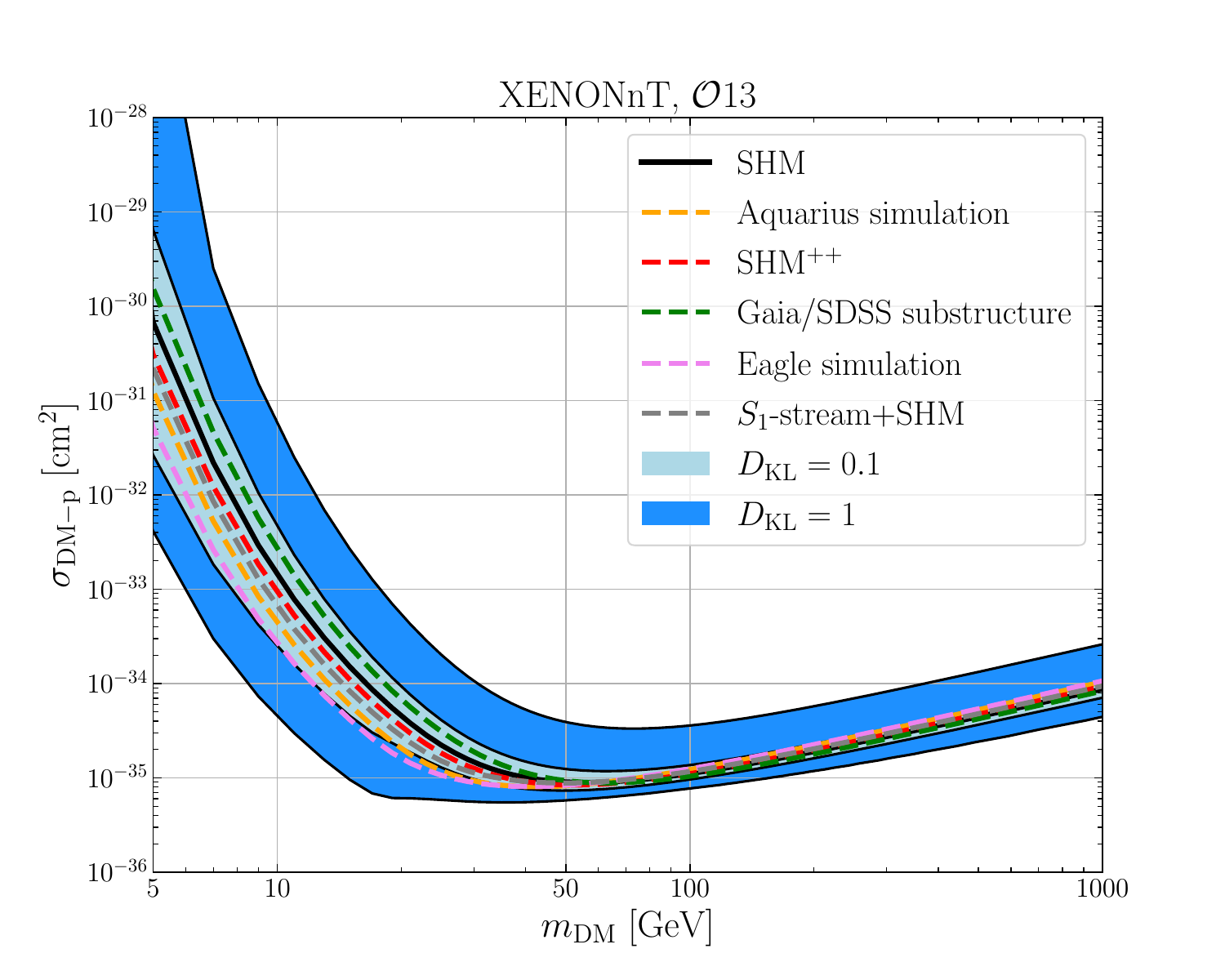}
\end{minipage}\hfill

\vspace{-2mm}

\begin{minipage}[H]{0.49\linewidth} % Adjust the width as desired
\includegraphics[width=\linewidth]{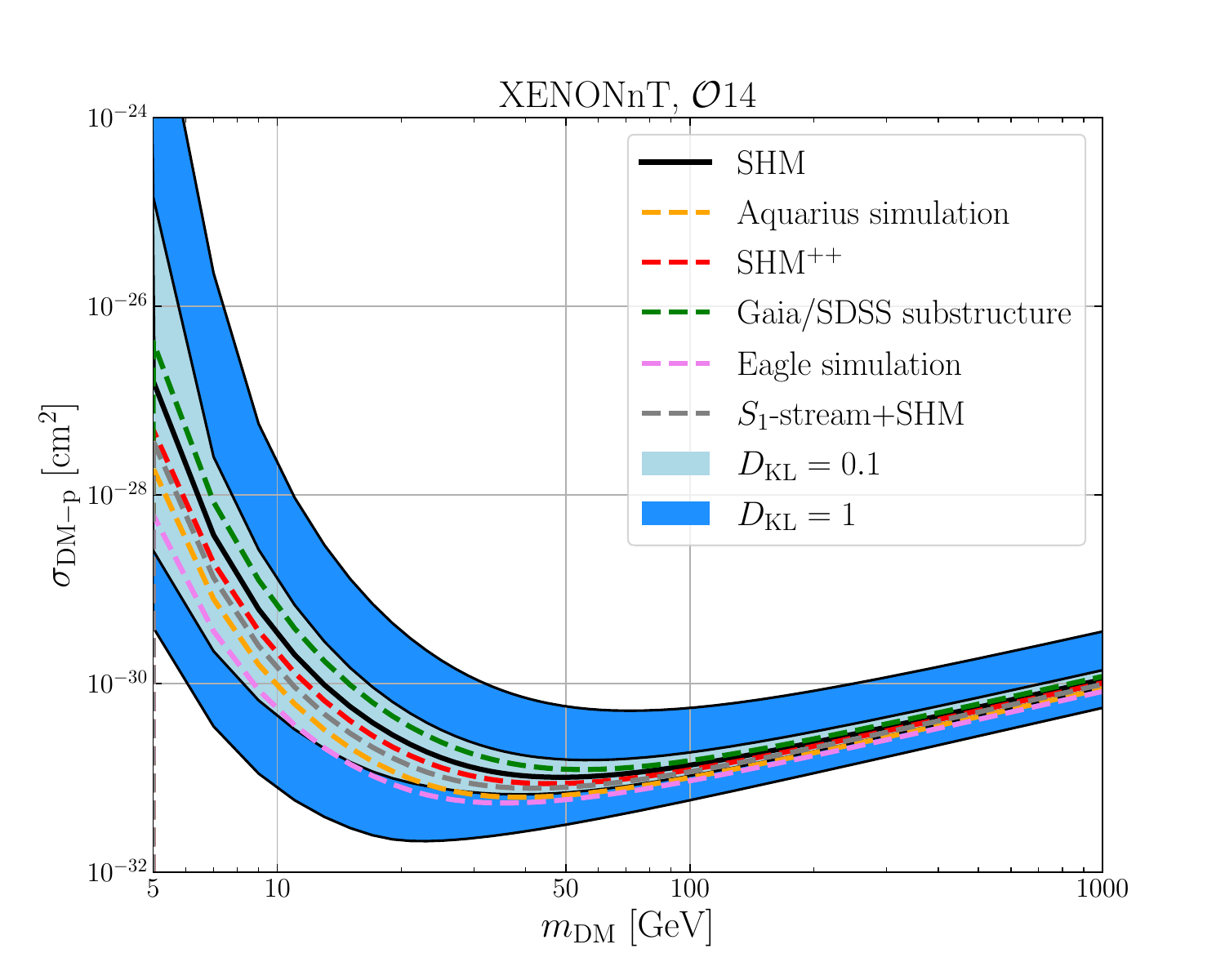}
\end{minipage}\hfill
\begin{minipage}[H]{0.49\linewidth} % Adjust the width as desired
\includegraphics[width=\linewidth]{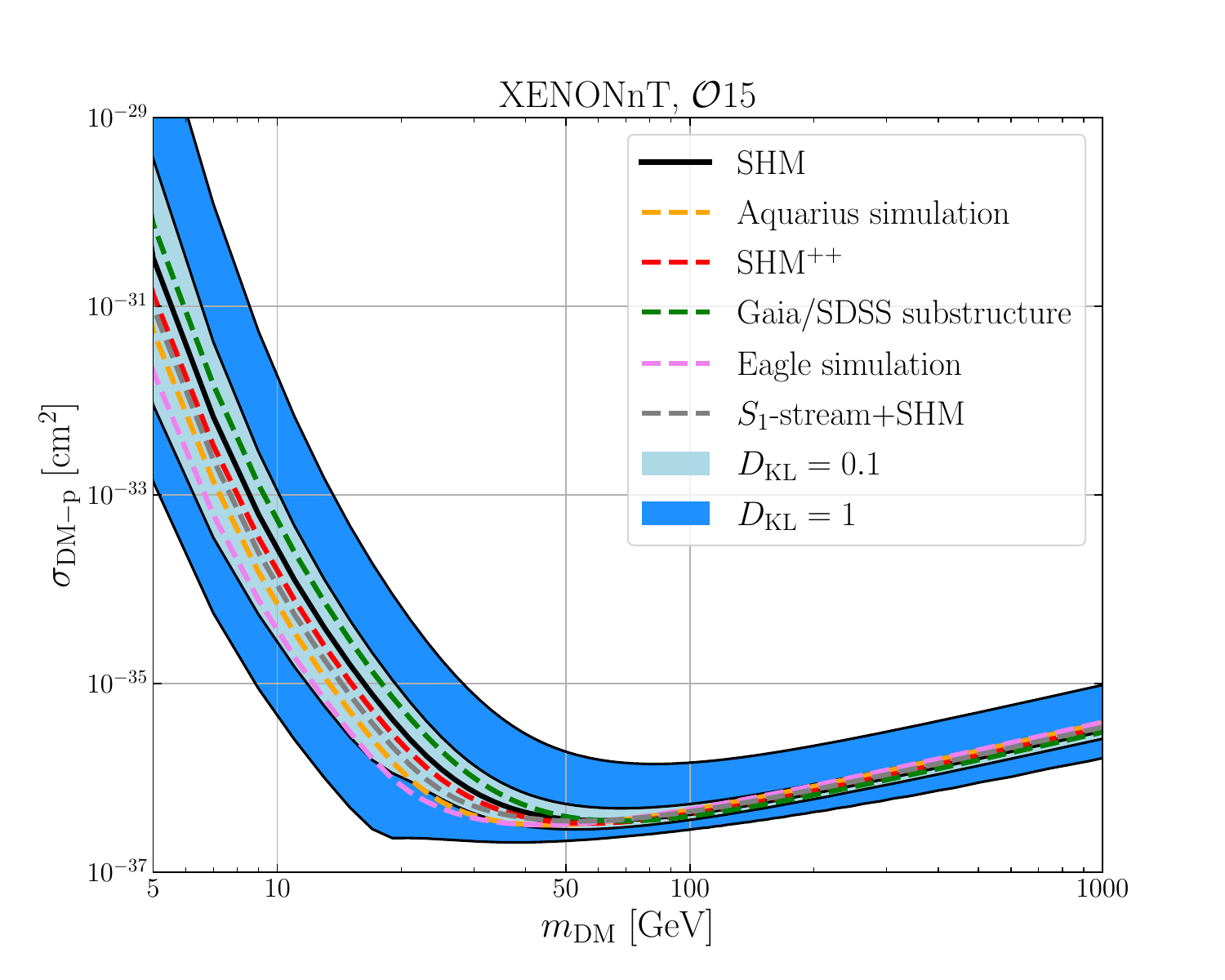}
\end{minipage}

\caption{90\%  C.L upper limits on the  dark matter-nucleon coupling from XENONnT, for different values of the KL-divergence between the Maxwell-Boltzmann distribution and the true velocity distribution. For comparison, we show the upper limits obtained from other velocity distributions motivated in the literature: Aquarius \cite{Vogelsberger:2008qb}, SHM$^{++}$ \cite{Evans:2018bqy}, Gaia/SDSS \cite{Necib:2018iwb}, Eagle \cite{Schaye:2014tpa}, and S1-stream+SHM \cite{OHare:2018trr}.}
\label{fig:xenon_KL_upperlimits_NREFT}
\end{figure}

\begin{figure}[H]
\begin{minipage}[H]{0.49\linewidth} % Adjust the width as desired
\includegraphics[width=\linewidth]{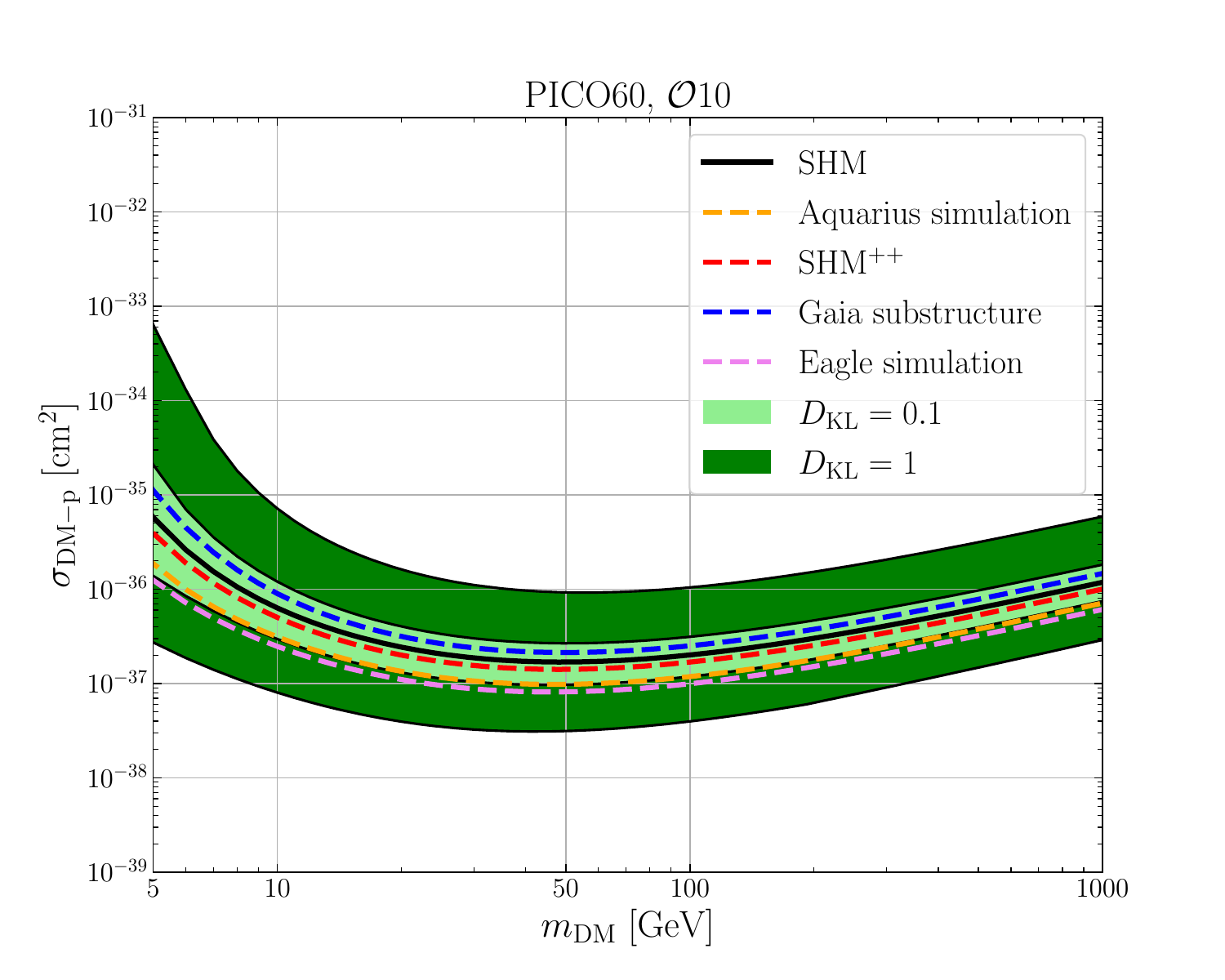}
\end{minipage}\hfill
\begin{minipage}[H]{0.49\linewidth} % Adjust the width as desired
\includegraphics[width=\linewidth]{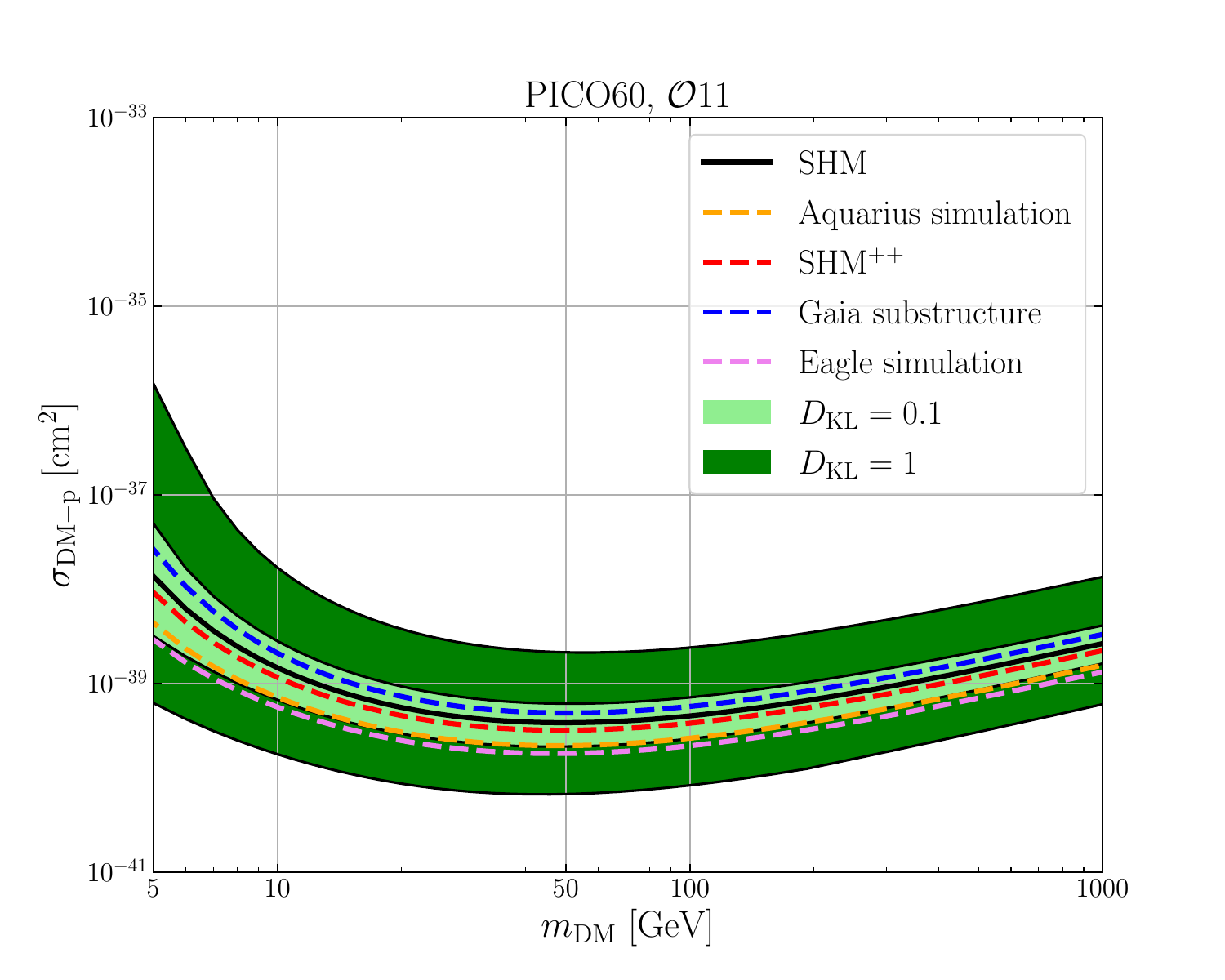}
\end{minipage}

\vspace{-2mm} % Adjust the vertical spacing as desired

\begin{minipage}[H]{0.49\linewidth} % Adjust the width as desired
\includegraphics[width=\linewidth]{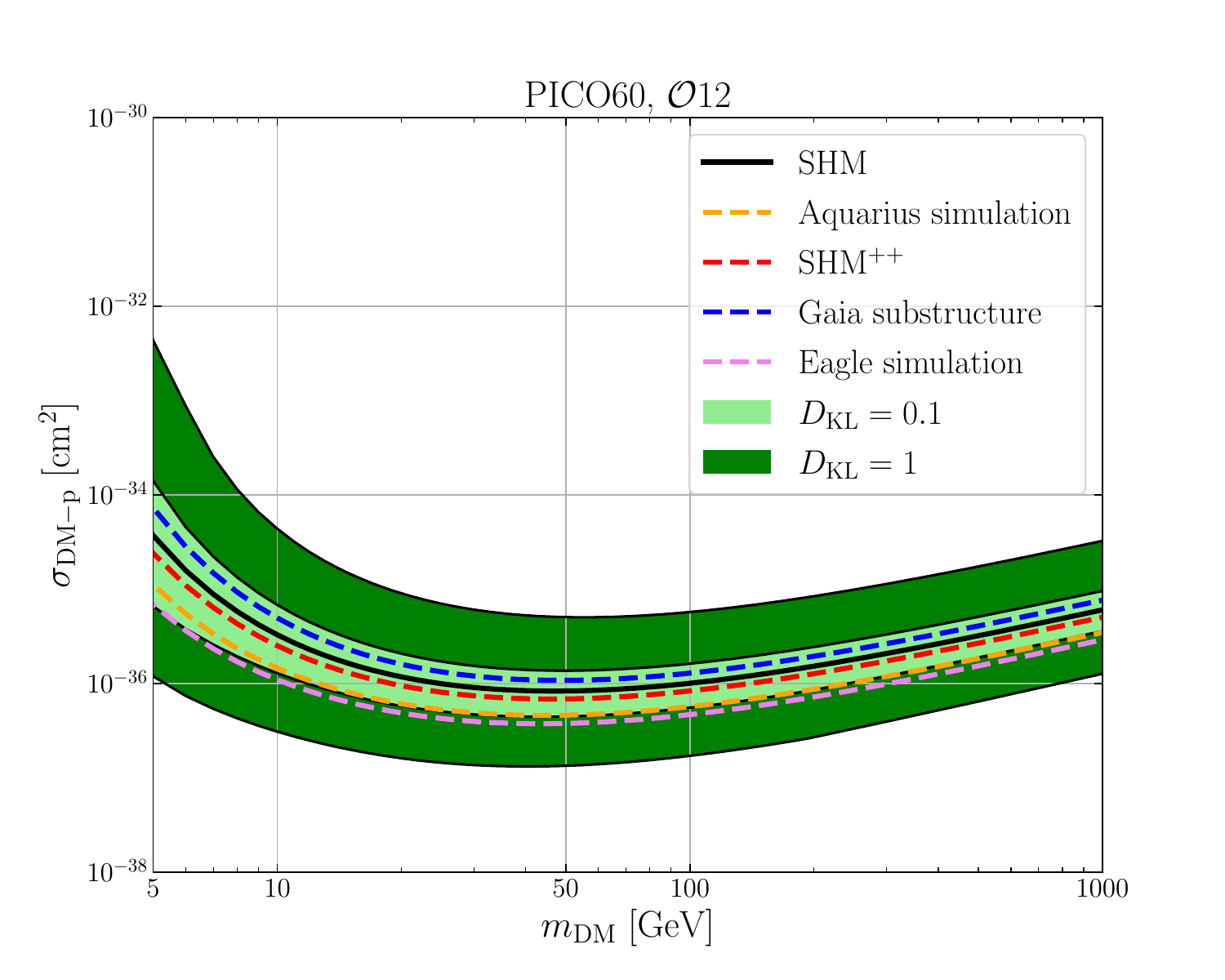}
\end{minipage}\hfill
\begin{minipage}[H]{0.49\linewidth} % Adjust the width as desired
\includegraphics[width=\linewidth]{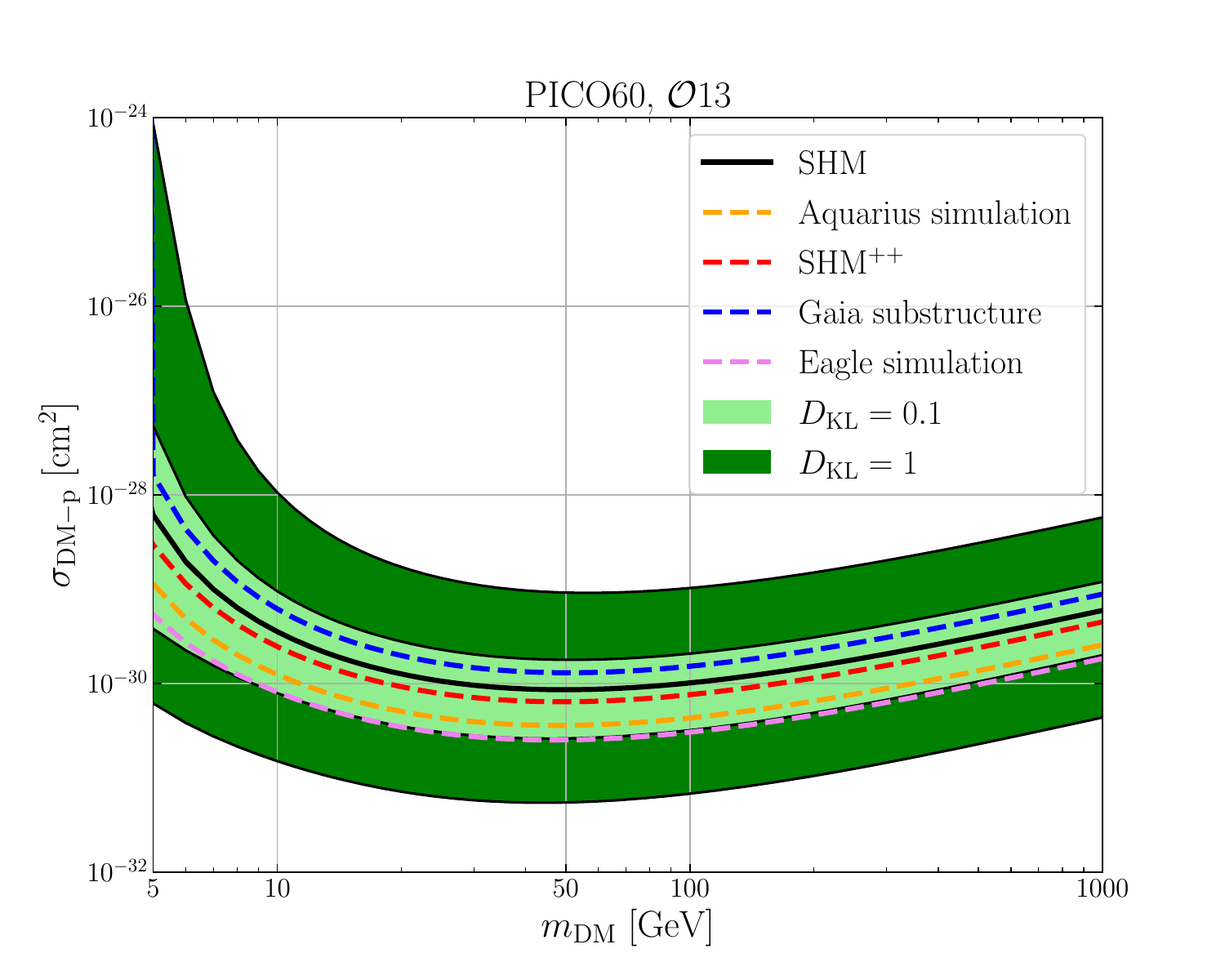}
\end{minipage}\hfill

\vspace{-2mm}

\begin{minipage}[H]{0.49\linewidth} % Adjust the width as desired
\includegraphics[width=\linewidth]{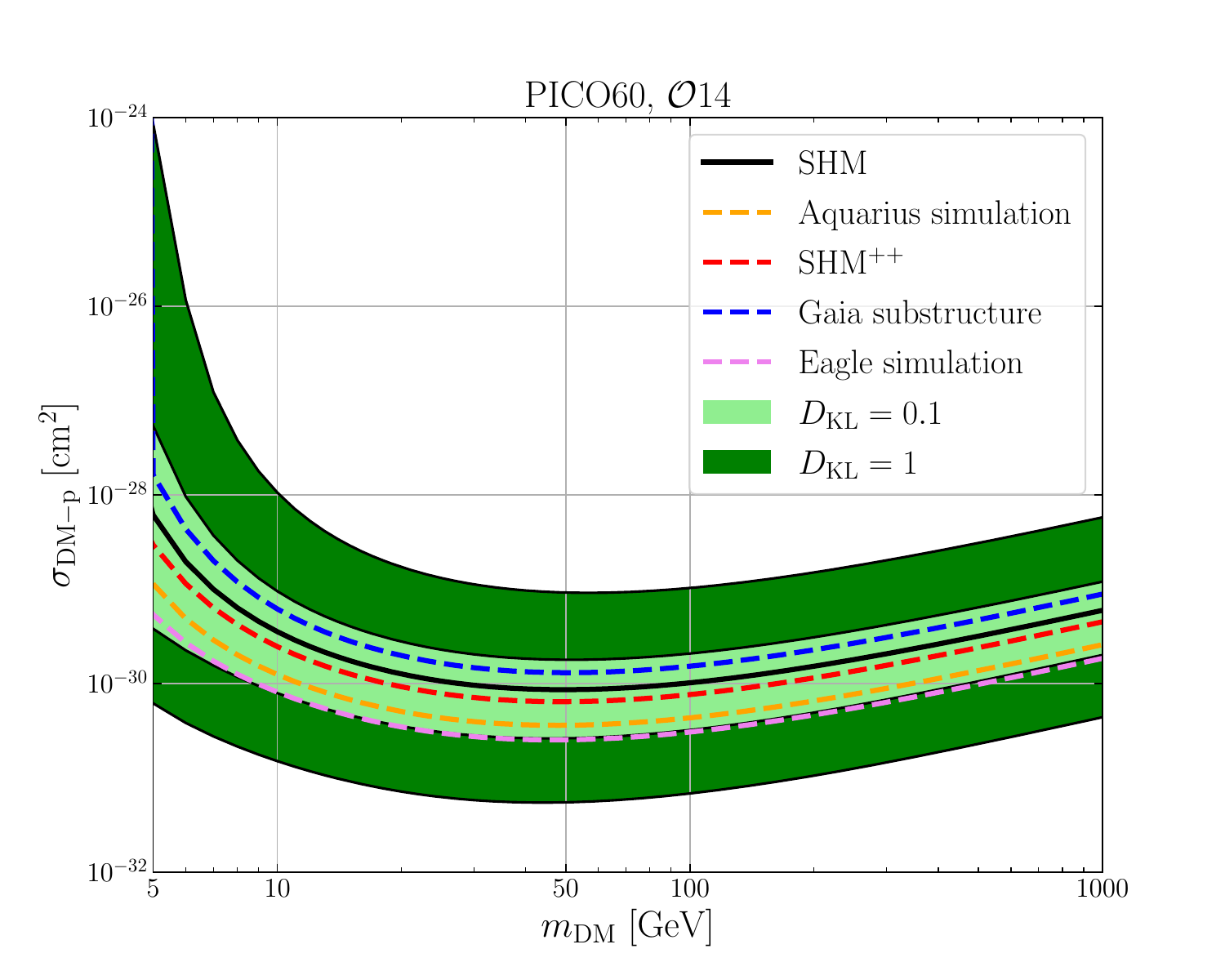}
\end{minipage}\hfill
\begin{minipage}[H]{0.49\linewidth} % Adjust the width as desired
\includegraphics[width=\linewidth]{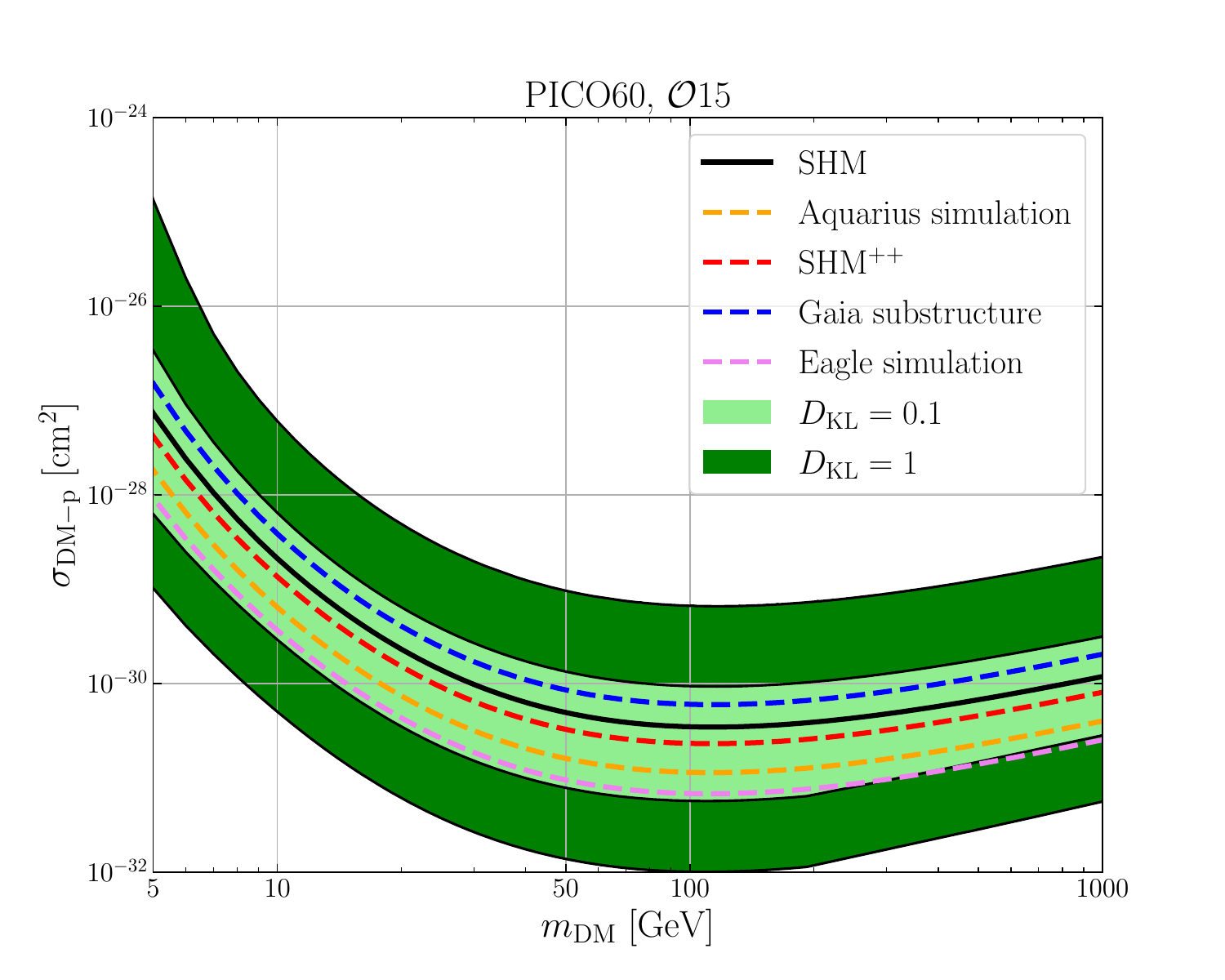}
\end{minipage}

\caption{90\%  C.L upper limits on the  dark matter-nucleon coupling from PICO60, for different values of the KL-divergence between the Maxwell-Boltzmann distribution and the true velocity distribution. For comparison, we show the upper limits from other velocity distributions motivated in the literature.}
\label{fig:pico60_KL_upperlimits_NREFT}
\end{figure}

\begin{figure}[H]
\centering
\setlength{\tabcolsep}{0pt} % Remove horizontal spacing between minipages
\begin{minipage}[H]{0.49\linewidth} % Adjust the width as desired
\includegraphics[width=\linewidth]{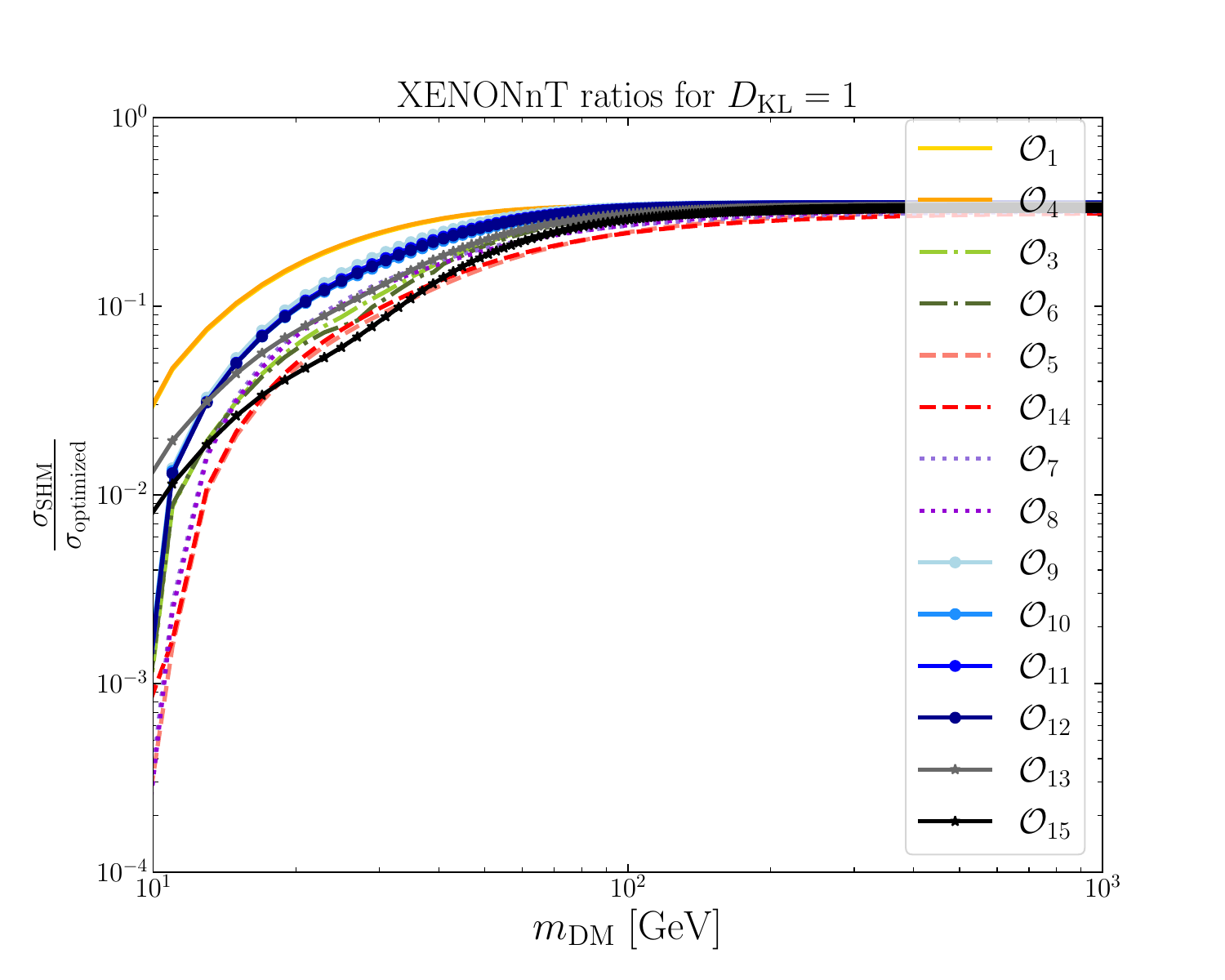}
\end{minipage}\hfill
\begin{minipage}[H]{0.49\linewidth} % Adjust the width as desired
\includegraphics[width=\linewidth]{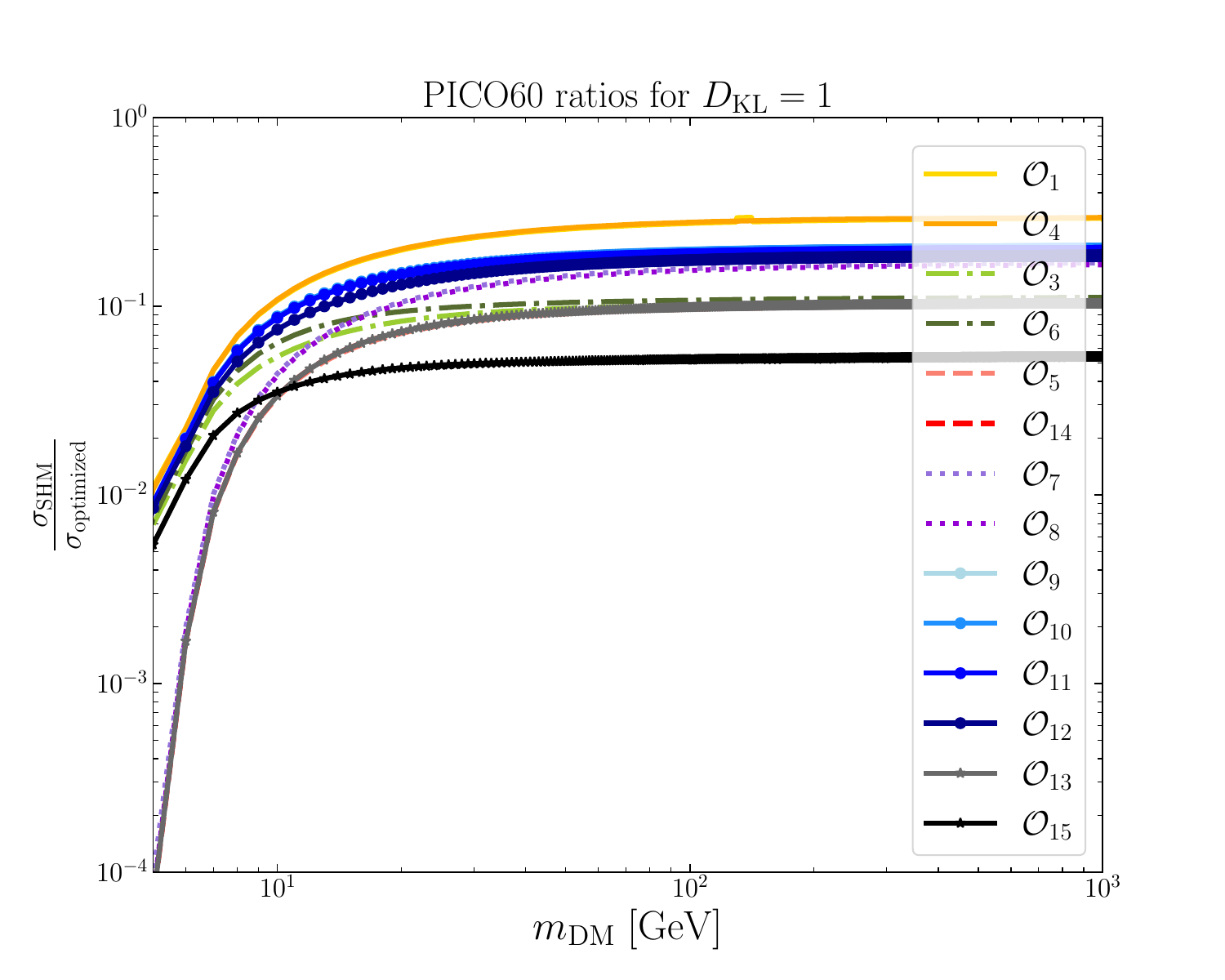}
\end{minipage}

\vspace{-2mm} % Adjust the vertical spacing as desired

\begin{minipage}[H]{0.49\linewidth} % Adjust the width as desired
\includegraphics[width=\linewidth]{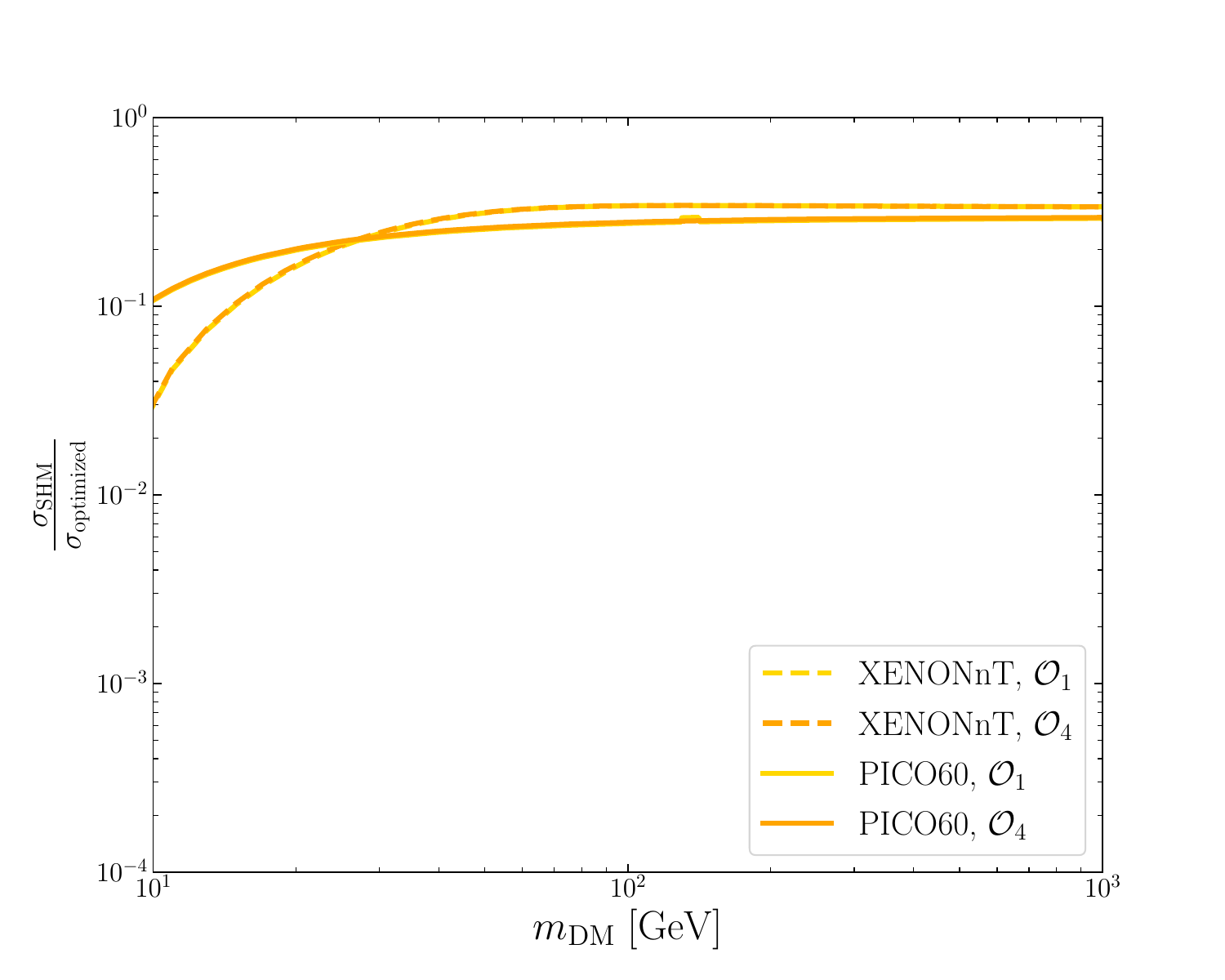}
\end{minipage}\hfill
\begin{minipage}[H]{0.49\linewidth} % Adjust the width as desired
\includegraphics[width=\linewidth]{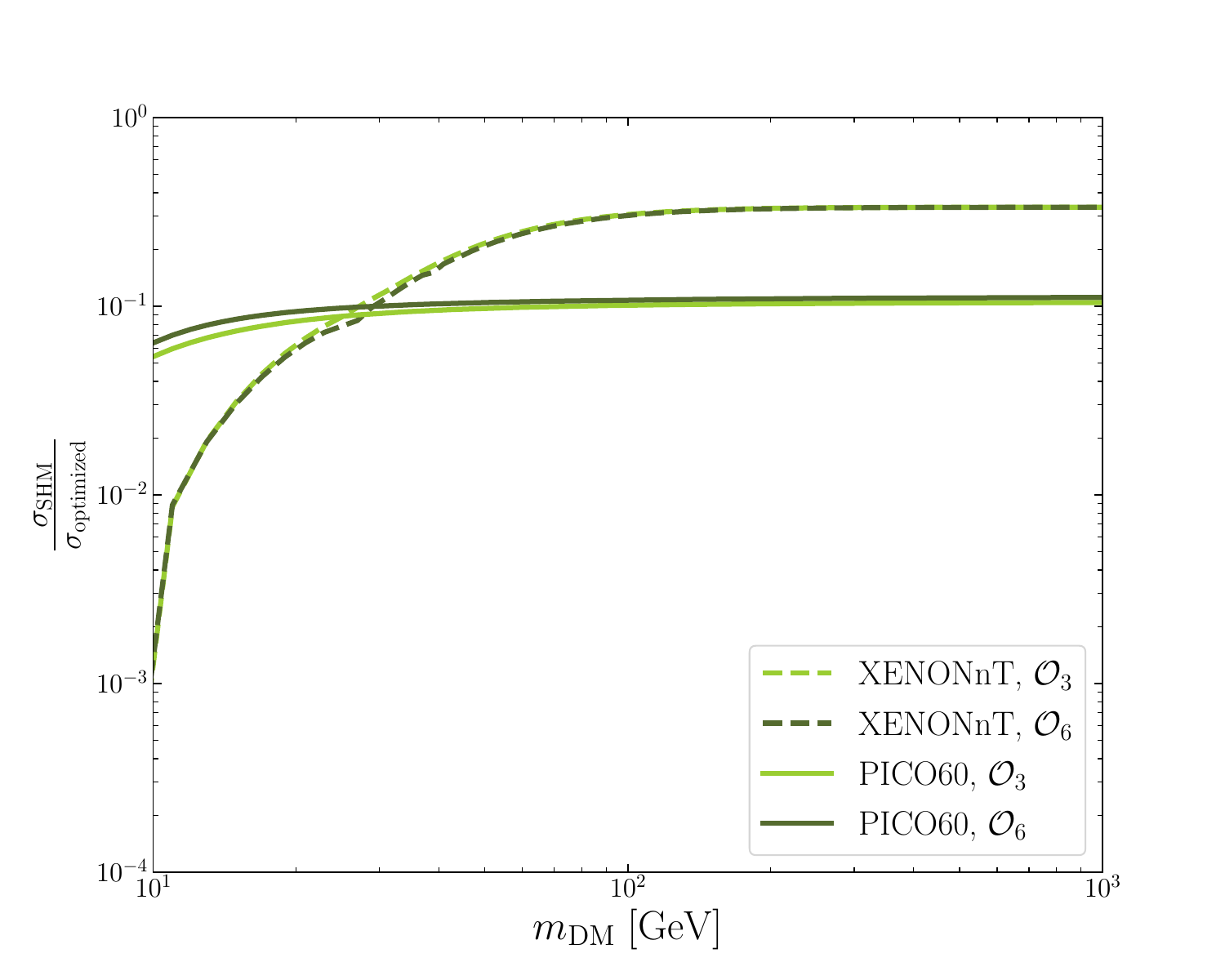}
\end{minipage}

\vspace{-2mm} % Adjust the vertical spacing as desired

\begin{minipage}[H]{0.49\linewidth} % Adjust the width as desired
\includegraphics[width=\linewidth]{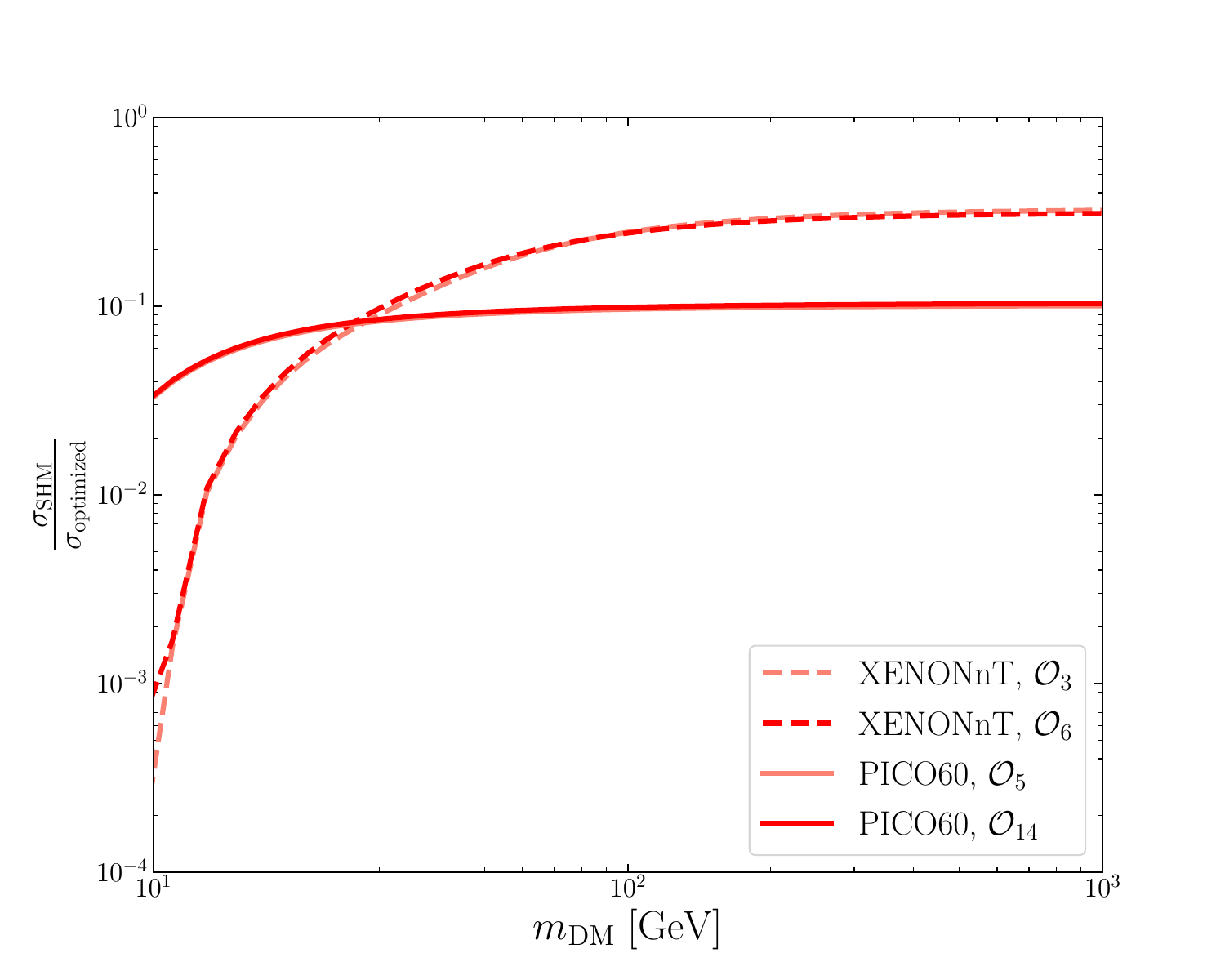}
\end{minipage}\hfill
\begin{minipage}[H]{0.49\linewidth} % Adjust the width as desired
\includegraphics[width=\linewidth]{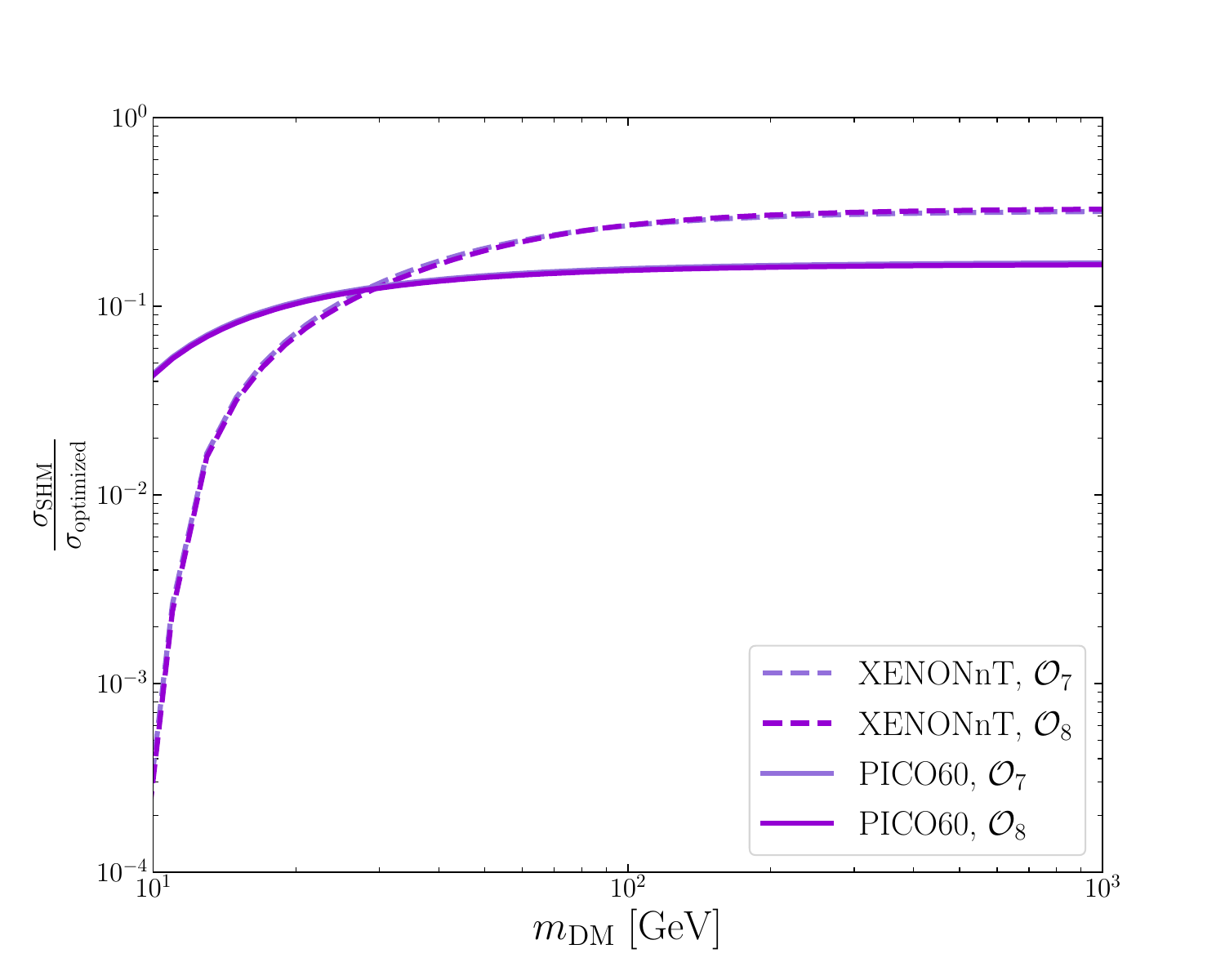}
\end{minipage}

\vspace{-2mm} % Adjust the vertical spacing as desired

\begin{minipage}[H]{0.49\linewidth} % Adjust the width as desired
\includegraphics[width=\linewidth]{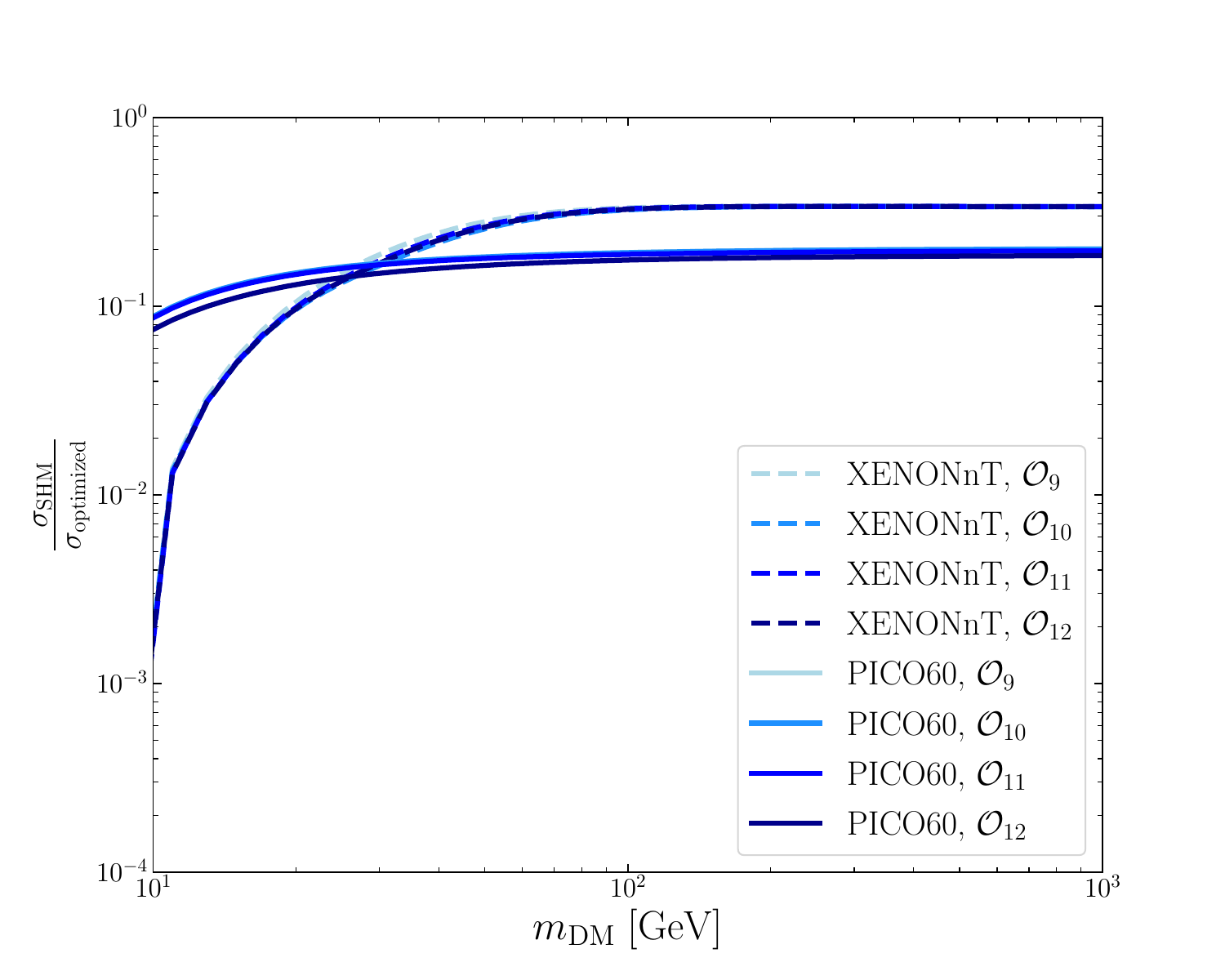}
\end{minipage}\hfill
\begin{minipage}[H]{0.49\linewidth} % Adjust the width as desired
\includegraphics[width=\linewidth]{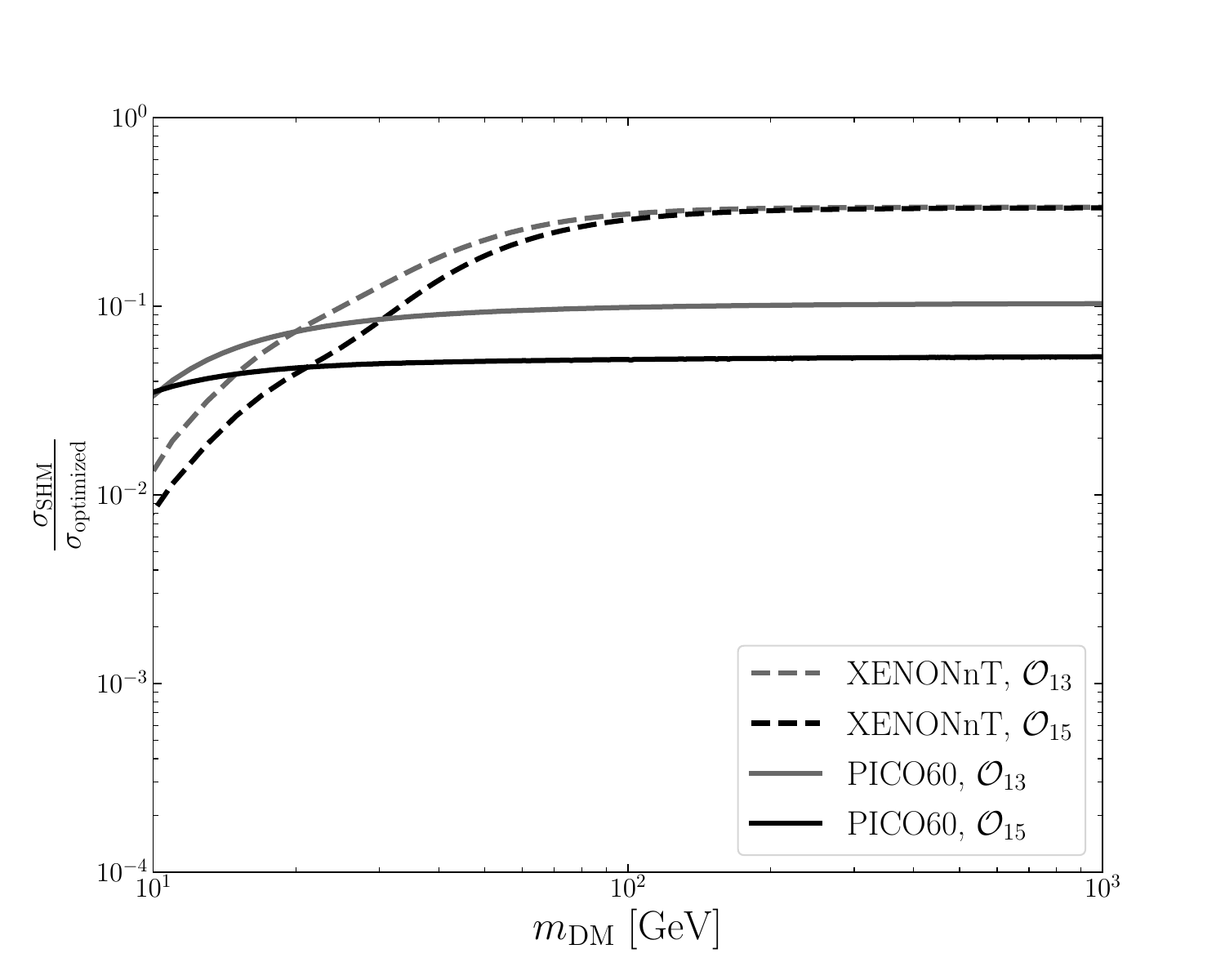}
\end{minipage}
\caption{Ratio of limits on the cross section in the SHM vs the KL-informed minima.}
\label{fig:ratios_KL_upperlimits_NREFT}
\end{figure}

In the following, we solve the optimization problem in Eq. \ref{eq:opt_problem_NREFT_VDF} finding the most conservative and most aggressive upper limits on the dark matter-nucleon scattering cross section from the XENONnT and PICO60 experiments, when considering deviations from the Maxwell-Boltzmann distribution parametrized by the KL-divergence with values $D_{\rm KL}=0.1$ and $ D_{\rm KL}=1$. The results are shown in Fig. \ref{fig:xenon_KL_upperlimits_NREFT} and Fig. \ref{fig:pico60_KL_upperlimits_NREFT} for several operators of the effective theory. Further, we also plot in Fig. \ref{fig:ratios_KL_upperlimits_NREFT} the ratio of upper limits in the scattering cross sections obtained with the Standard Halo Model (SHM), and when solving this optimization problem, for all operators of the effective theory. We group the operators depending on their velocity-dependencies according to the classification delineated in Section \ref{sec:methodology}.

We note that, as expected, the impact of astrophysical uncertainties is larger for those operators with a stronger velocity dependence. For instance, for velocity-independent operators such as $\mathcal{O}_1$ and $\mathcal{O}_4$ the uncertainties near threshold are small, of about one order of magnitude for $D_{\rm KL}=0.1$. On the other hand, for velocity-dependent operators such as $\mathcal{O}_3$ the uncertainties are larger near threshold, of about two orders of magnitude. For operators with a cubic velocity-weighting suh as $\mathcal{O}_5$ or $\mathcal{O}_8$, the uncertainties are even larger, of about three orders of magnitude near threshold. In addition, we find that the operator $\mathcal{O}_{14}$ displays particularly large astrophysical uncertainties, approaching four orders of magnitude in the near-threshold regime.

In Fig. \ref{fig:ratios_KL_upperlimits_NREFT} it can be clearly noticed that operators with the same velocity and momentum-dependence yield similar ratios on the the SHM-derived upper limit versus informed-theoretic derived ones in this work.  At high masses $m_{\rm DM} \gtrsim 20$ GeV, the impact of astrophysical uncertainties is very similar for all operators of the effective theory, of about a factor of three for $D_{\rm KL}=0.1$ in XENONnT, and of a factor of ten in PICO60.

%%%%%%%%%%%%%%%%%%%%%%%%%%%%%%%%%%%%
%%%%%%%%%%%%%%%%%%%%%%%%%%%%%%%%%%%%%%%%%
\begin{figure}[H]
\centering
\setlength{\tabcolsep}{0pt} % Remove horizontal spacing between minipages
\begin{minipage}[H]{0.49\linewidth} % Adjust the width as desired
\includegraphics[width=\linewidth]{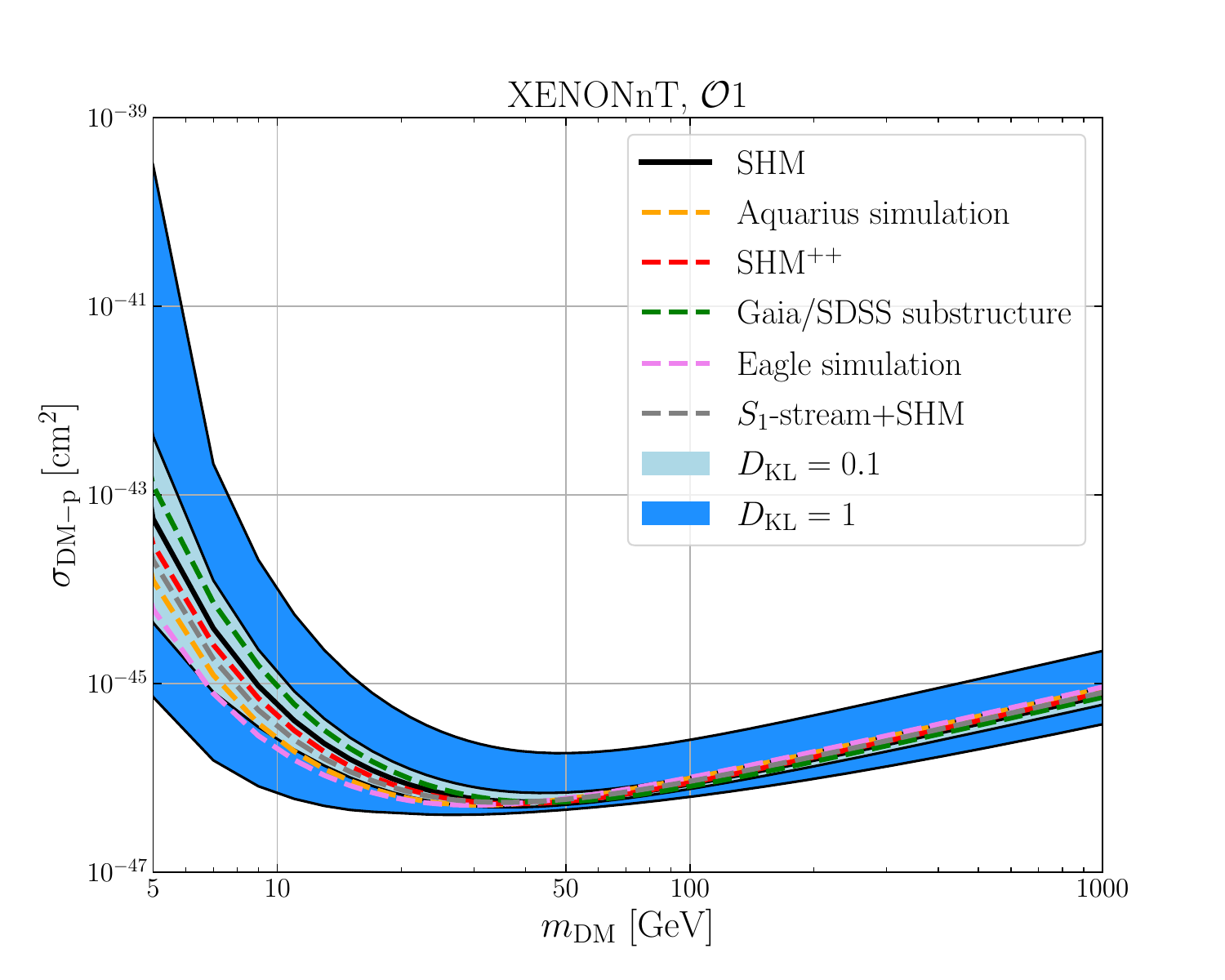}
\end{minipage}\hfill
\begin{minipage}[H]{0.49\linewidth} % Adjust the width as desired
\includegraphics[width=\linewidth]{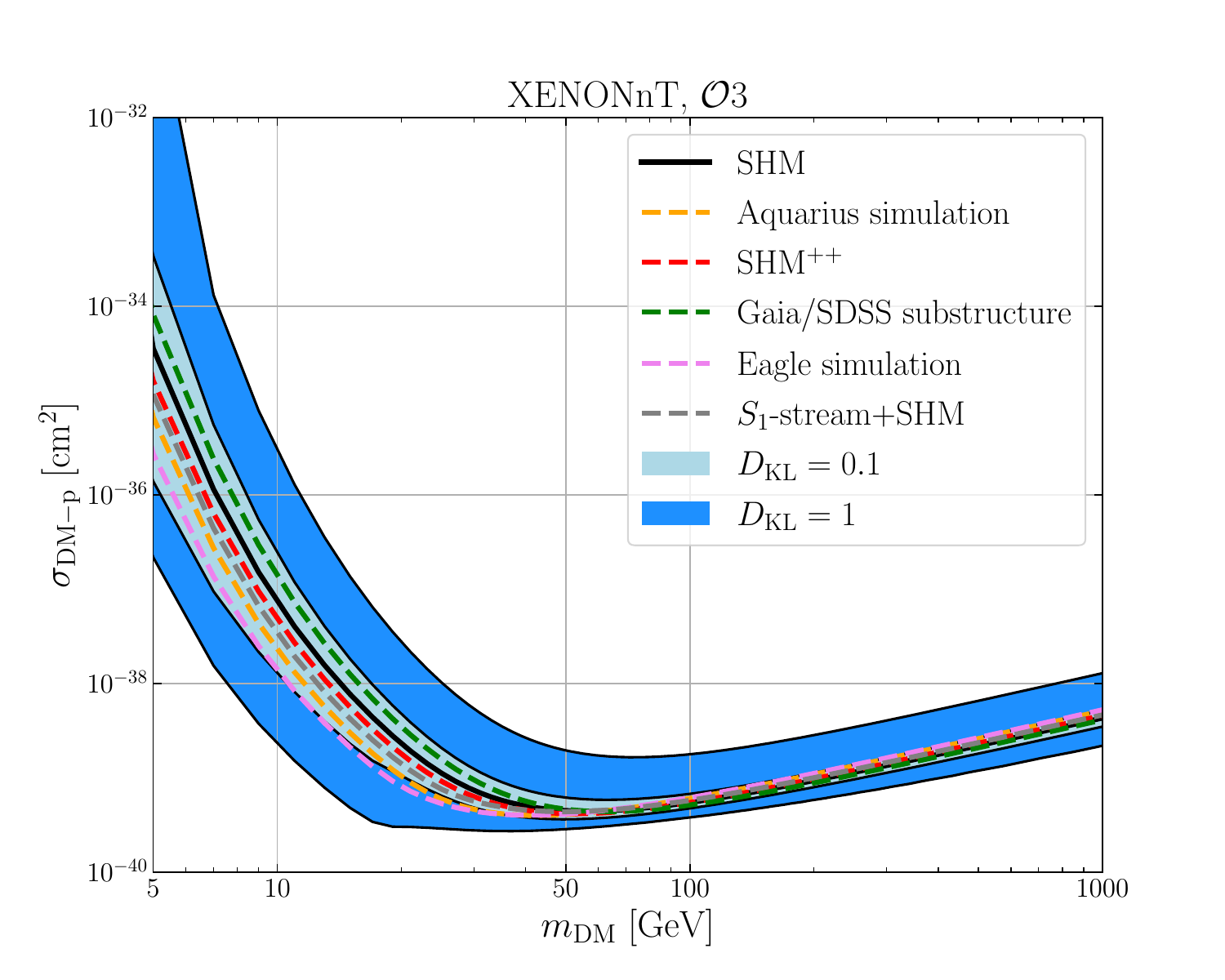}
\end{minipage}

%\vspace{-2mm} % Adjust the vertical spacing as desired

\begin{minipage}[H]{0.49\linewidth} % Adjust the width as desired
\includegraphics[width=\linewidth]{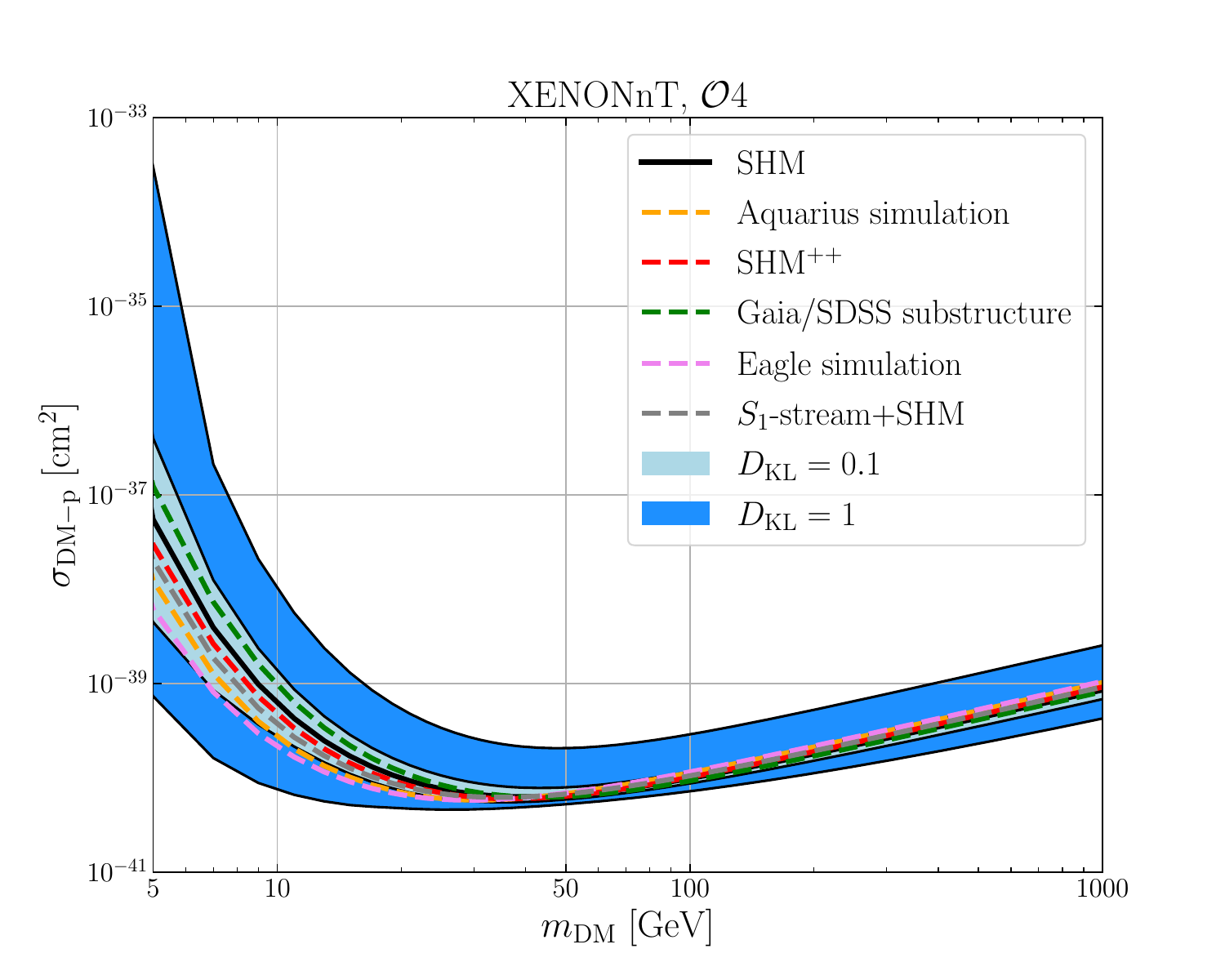}
\end{minipage}\hfill
\begin{minipage}[H]{0.49\linewidth} % Adjust the width as desired
\includegraphics[width=\linewidth]{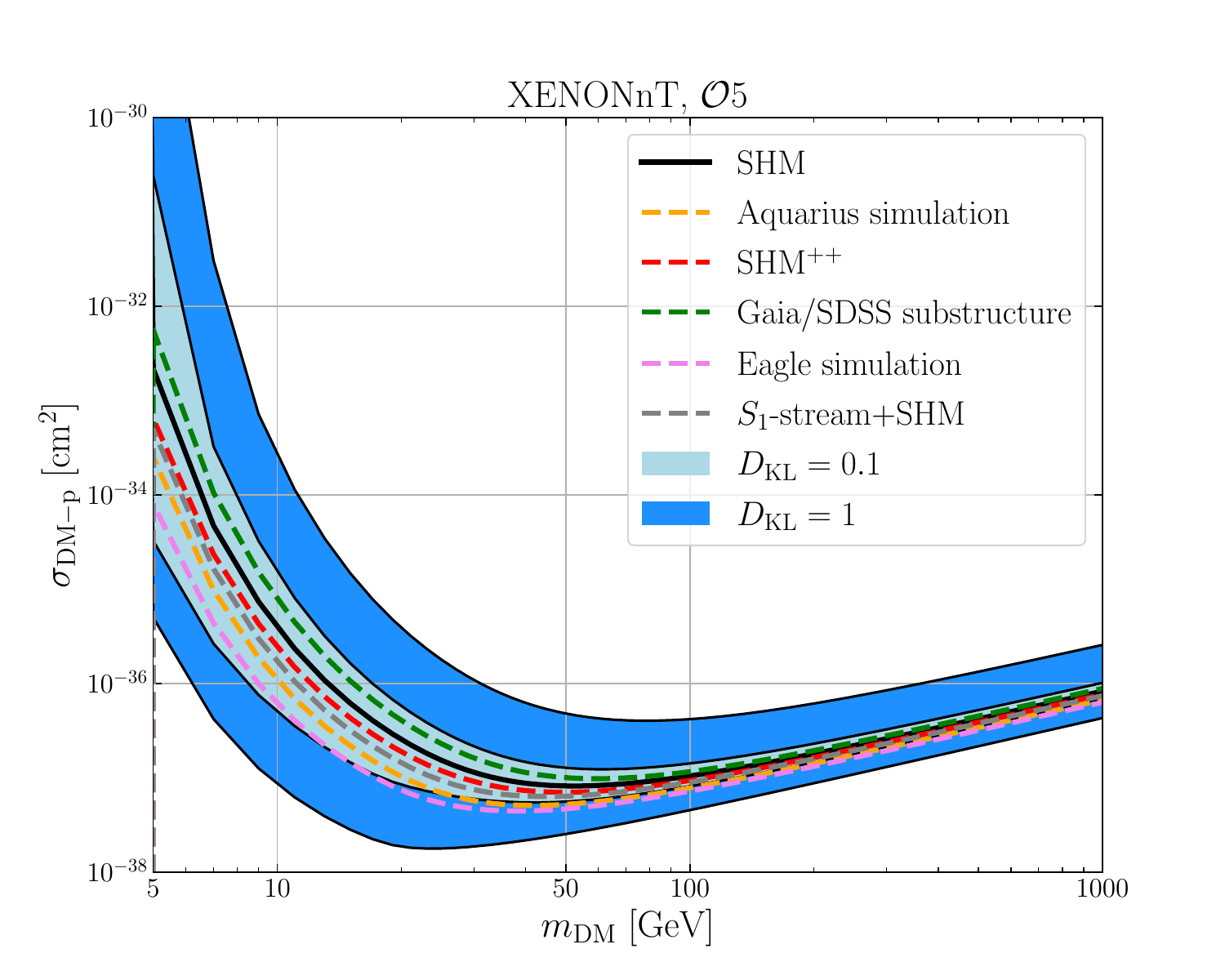}
\end{minipage}

%\vspace{-2mm} % Adjust the vertical spacing as desired

%\begin{minipage}[H]{0.49\linewidth} % Adjust the width as desired
%\includegraphics[width=\linewidth]{xenon_O6_astrovdf_5GeV.pdf}
%\end{minipage}\hfill
%\begin{minipage}[H]{0.49\linewidth} % Adjust the width as desired
%\includegraphics[width=\linewidth]{xenon_O7_astrovdf_5GeV.pdf}
%\end{minipage}

\vspace{-2mm} % Adjust the vertical spacing as desired

\begin{minipage}[H]{0.49\linewidth} % Adjust the width as desired
\includegraphics[width=\linewidth]{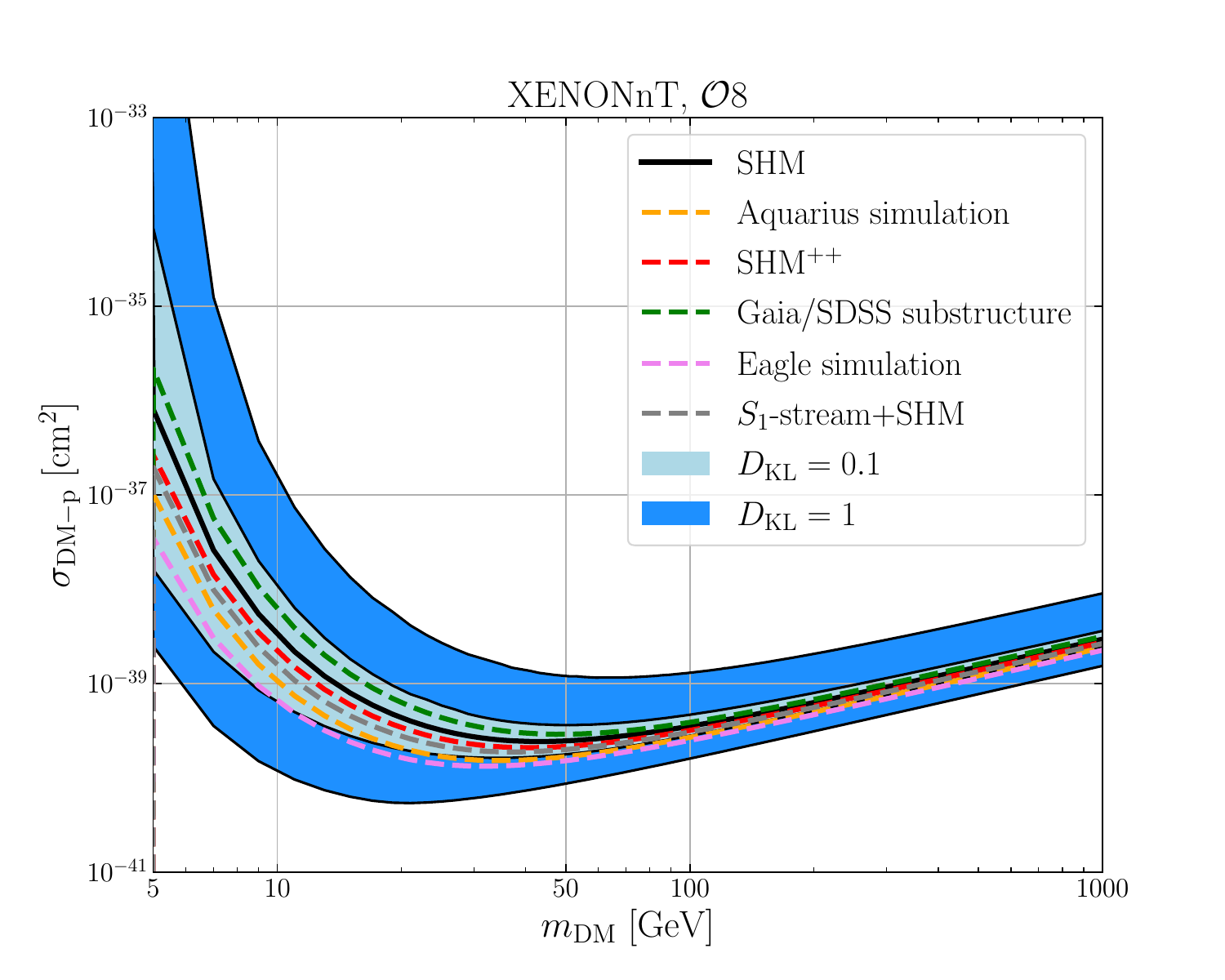}
\end{minipage}\hfill
\begin{minipage}[H]{0.49\linewidth} % Adjust the width as desired
\includegraphics[width=\linewidth]{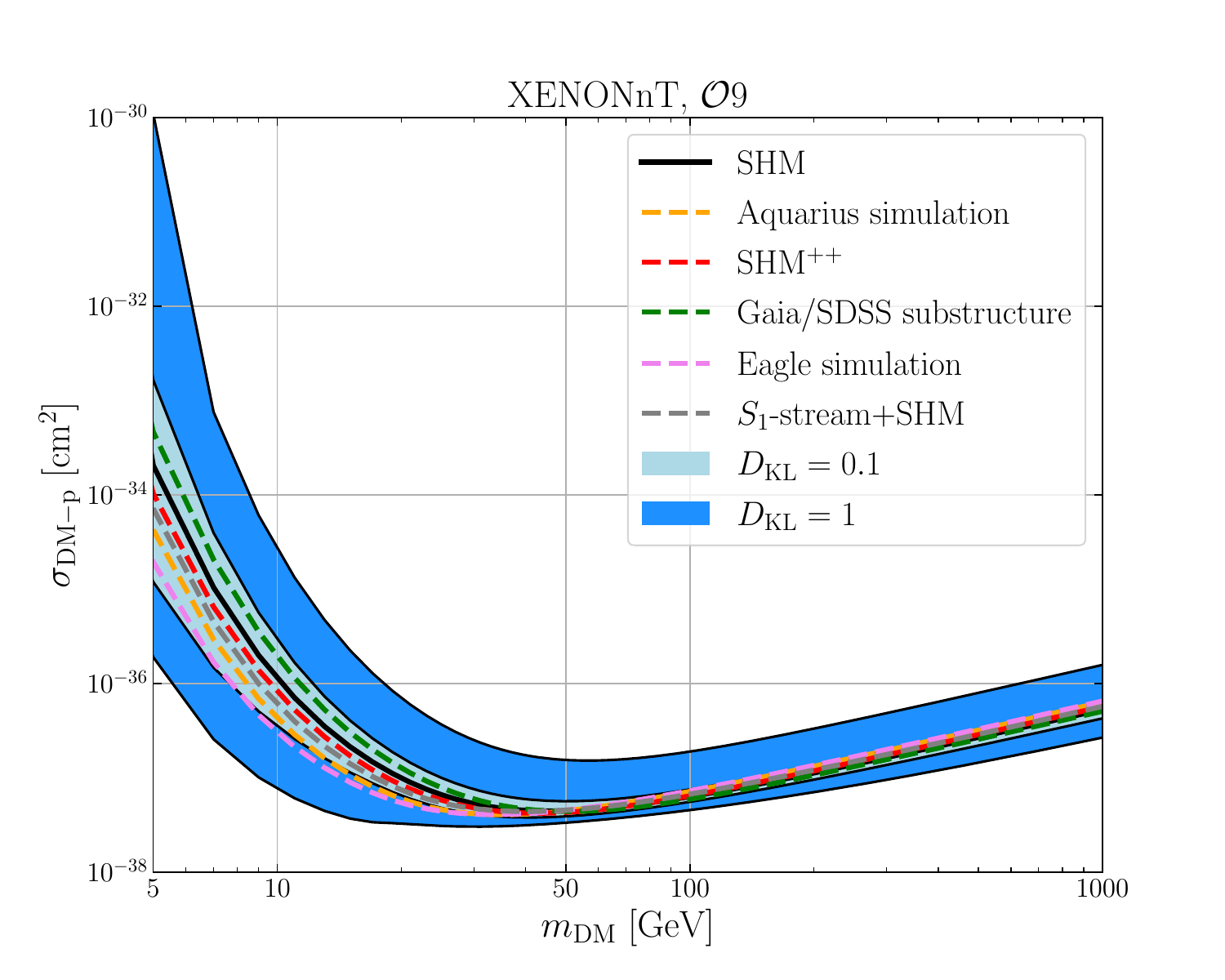}
\end{minipage}
\end{figure}

%#######################################
\begin{figure}[H]
\centering
\setlength{\tabcolsep}{0pt} % Remove horizontal spacing between minipages
\begin{minipage}[H]{0.49\linewidth} % Adjust the width as desired
\includegraphics[width=\linewidth]{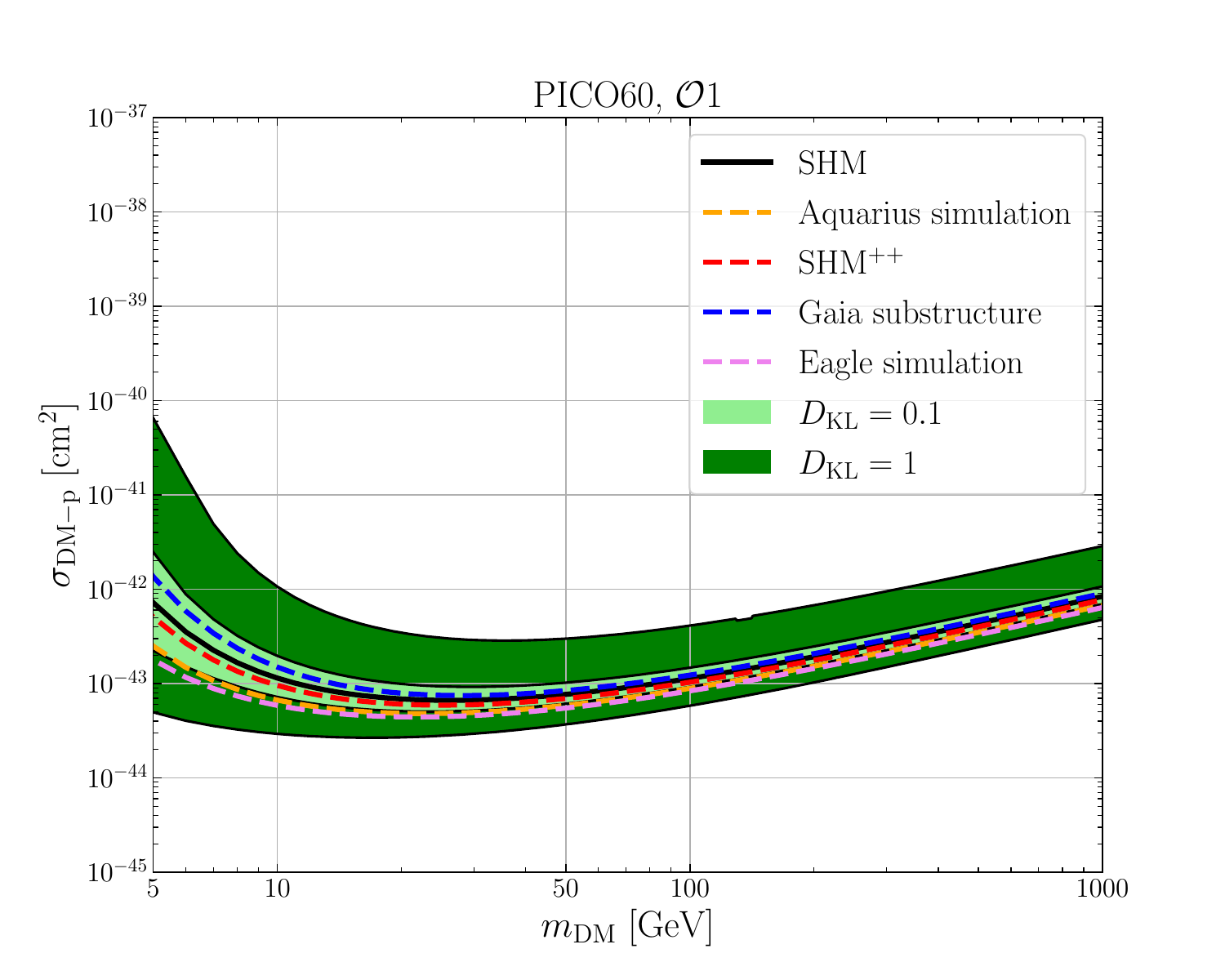}
\end{minipage}\hfill
\begin{minipage}[H]{0.49\linewidth} % Adjust the width as desired
\includegraphics[width=\linewidth]{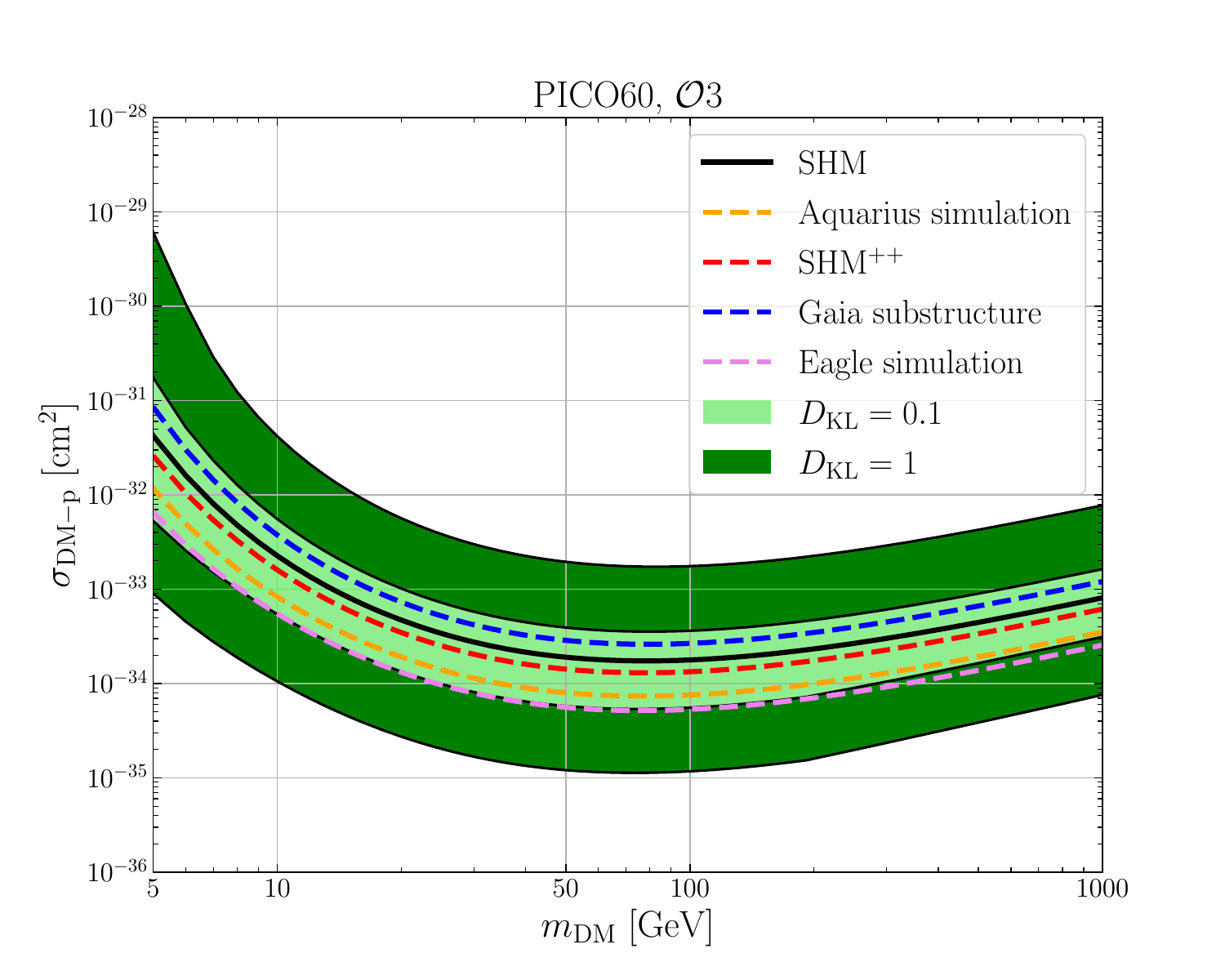}
\end{minipage}

\vspace{-2mm} % Adjust the vertical spacing as desired

\begin{minipage}[H]{0.49\linewidth} % Adjust the width as desired
\includegraphics[width=\linewidth]{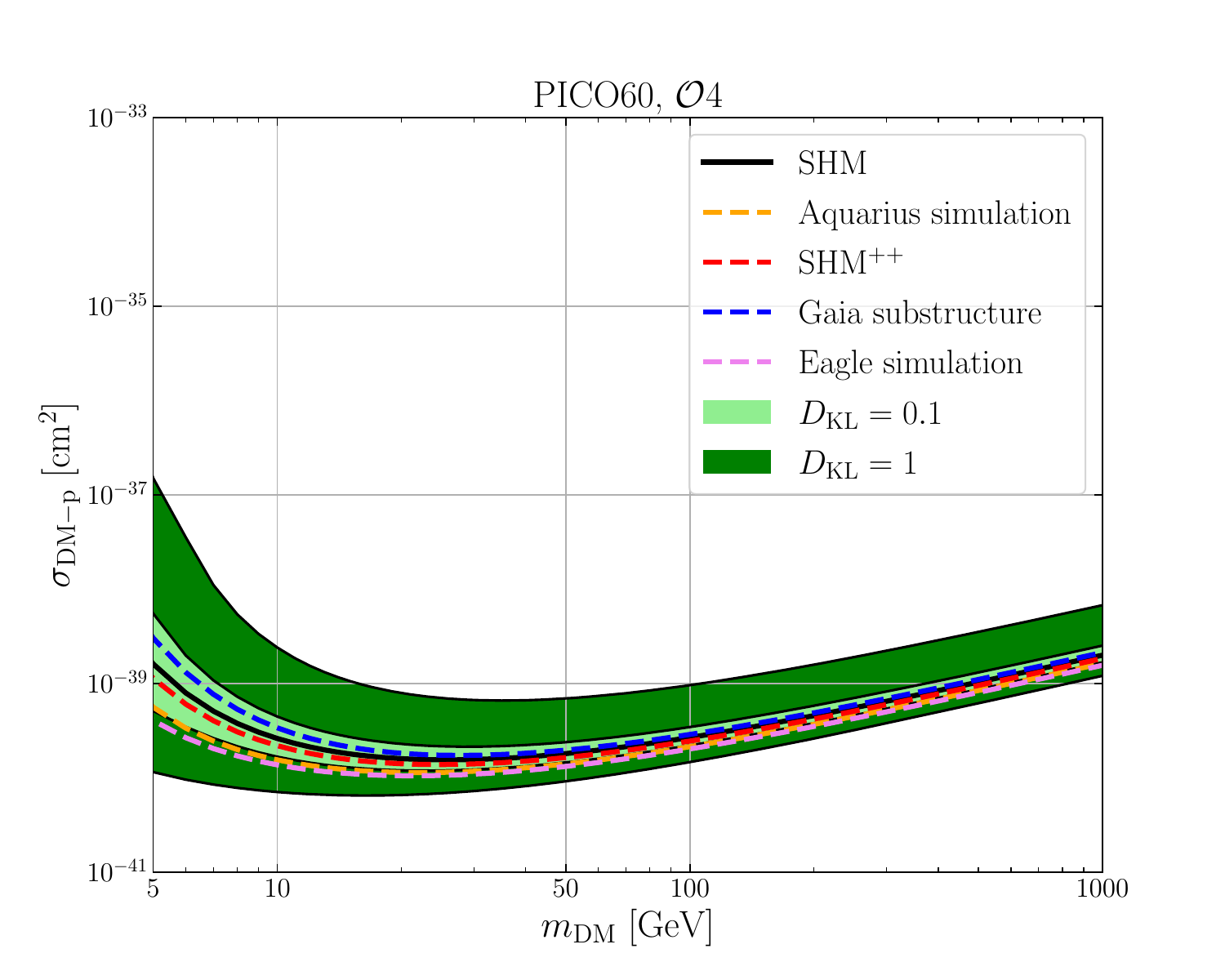}
\end{minipage}\hfill
\begin{minipage}[H]{0.49\linewidth} % Adjust the width as desired
\includegraphics[width=\linewidth]{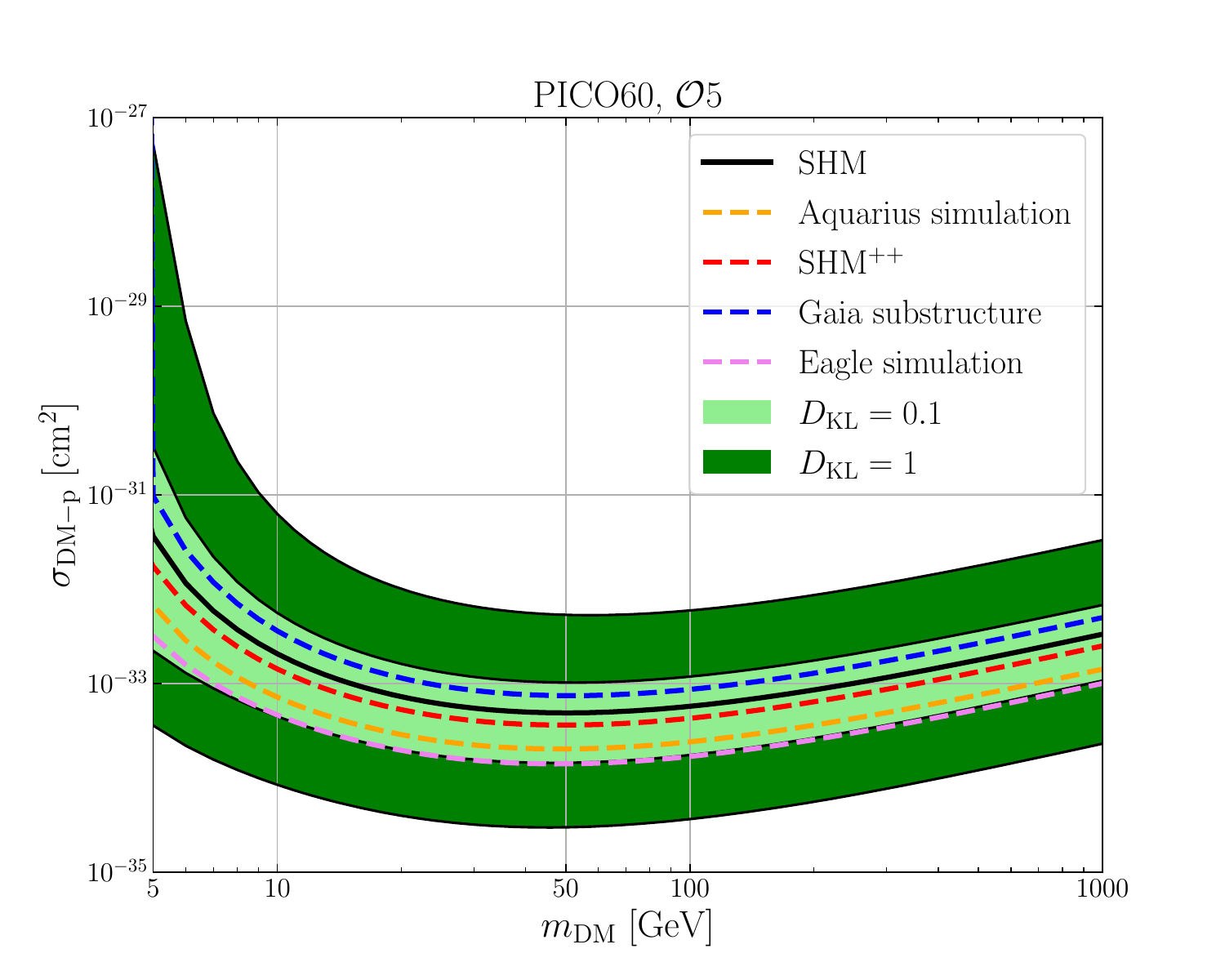}
\end{minipage}

\vspace{-2mm} % Adjust the vertical spacing as desired

\begin{minipage}[H]{0.49\linewidth} % Adjust the width as desired
\includegraphics[width=\linewidth]{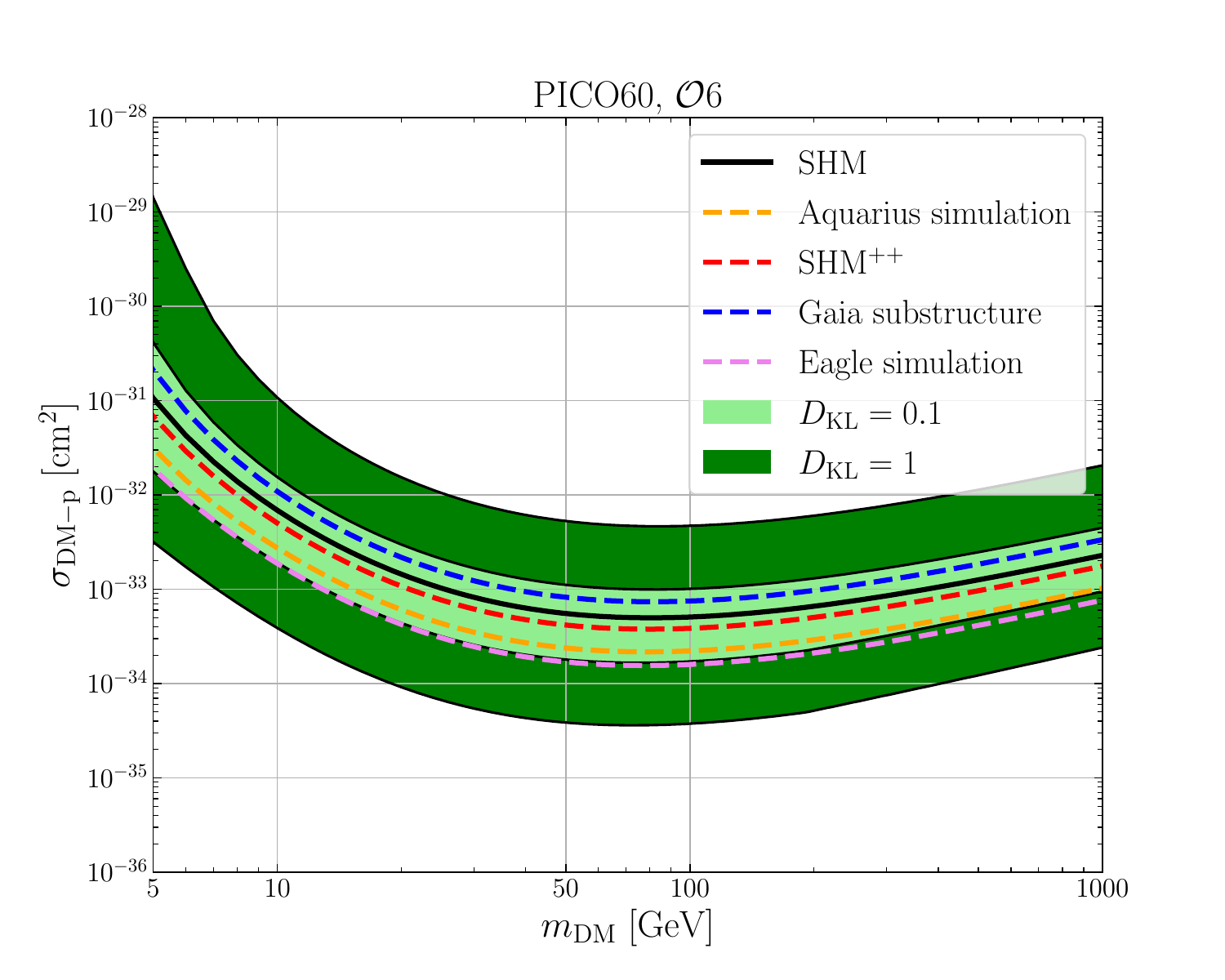}
\end{minipage}\hfill
\begin{minipage}[H]{0.49\linewidth} % Adjust the width as desired
\includegraphics[width=\linewidth]{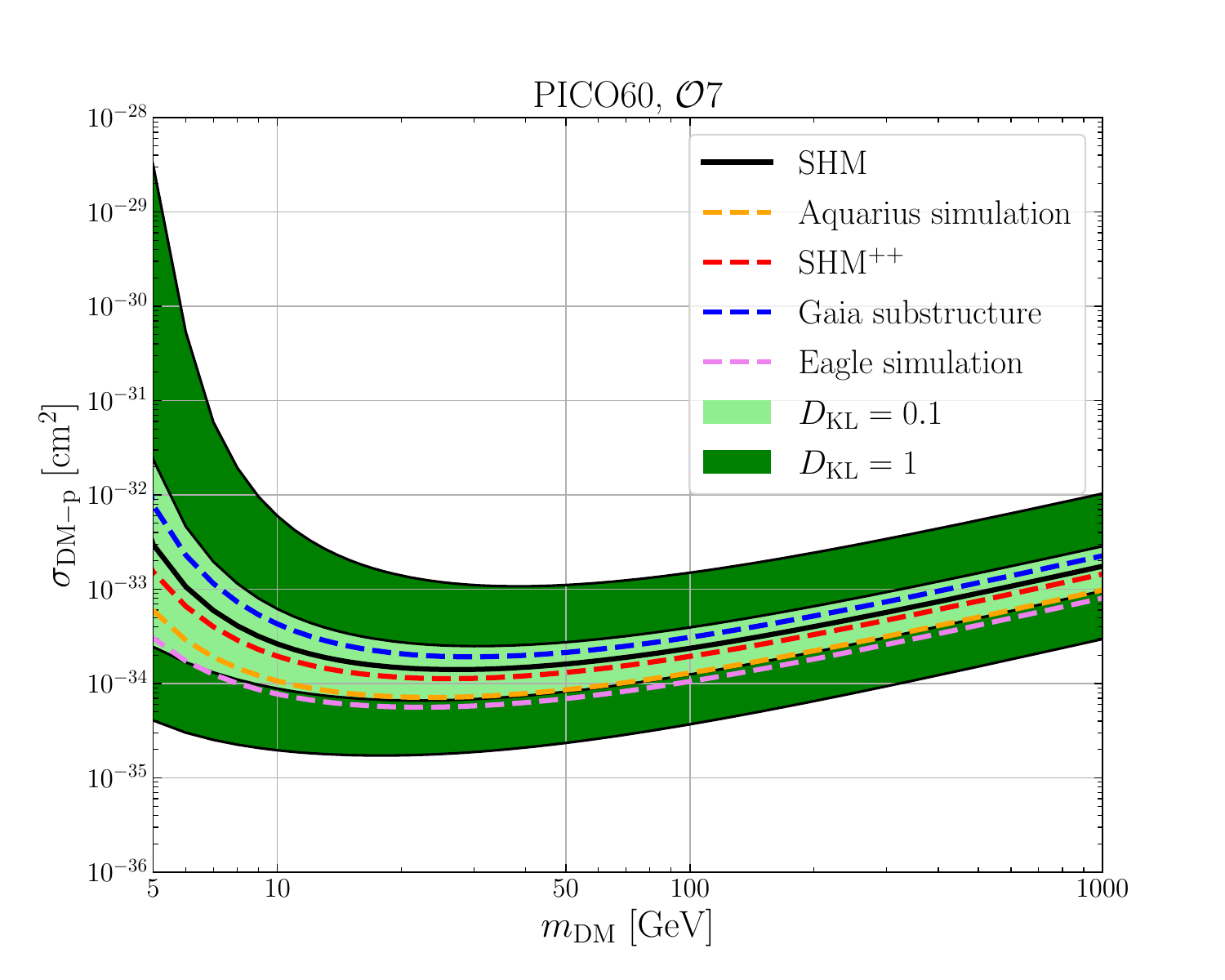}
\end{minipage}

\vspace{-2mm} % Adjust the vertical spacing as desired

\begin{minipage}[H]{0.49\linewidth} % Adjust the width as desired
\includegraphics[width=\linewidth]{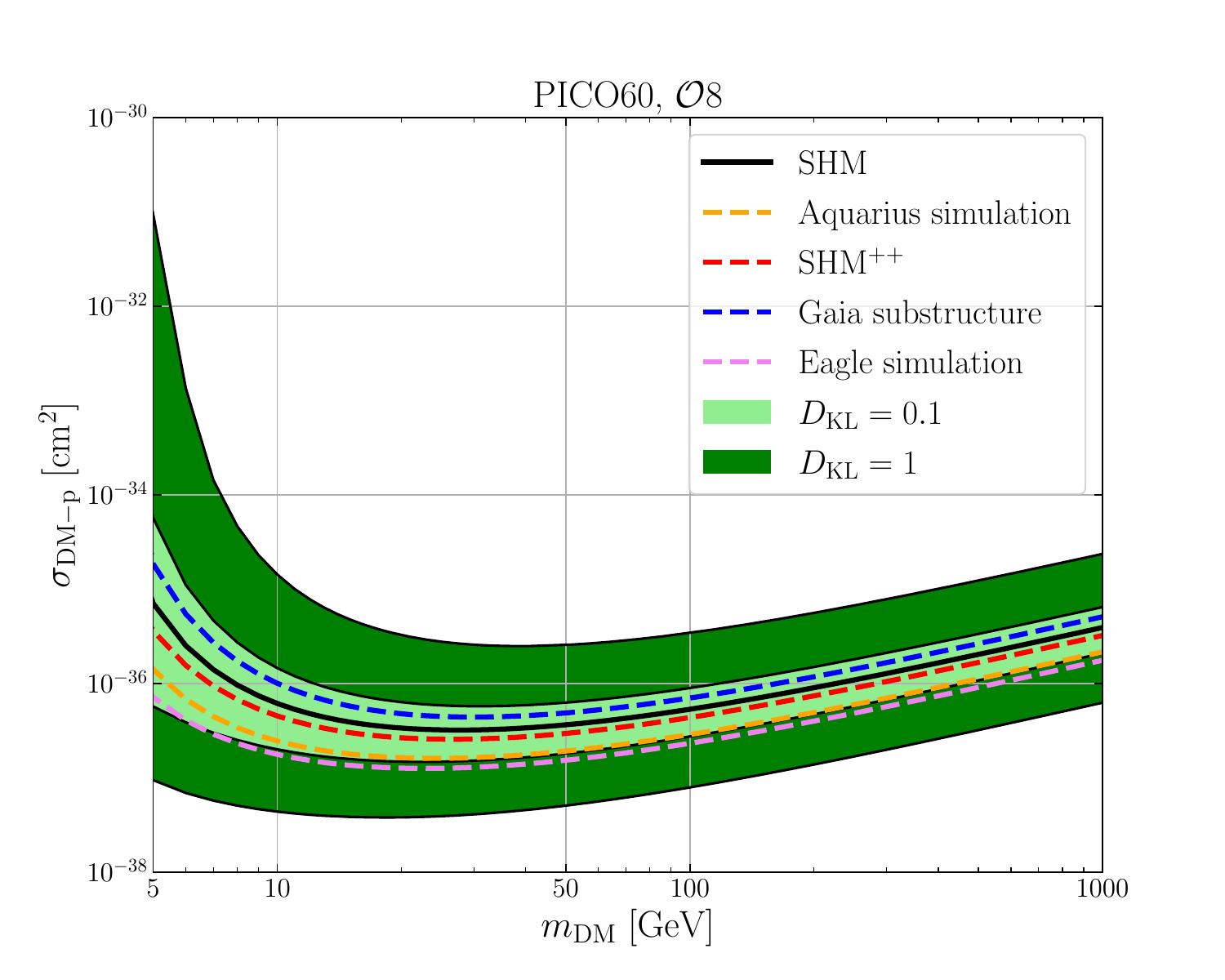}
\end{minipage}\hfill
\begin{minipage}[H]{0.49\linewidth} % Adjust the width as desired
\includegraphics[width=\linewidth]{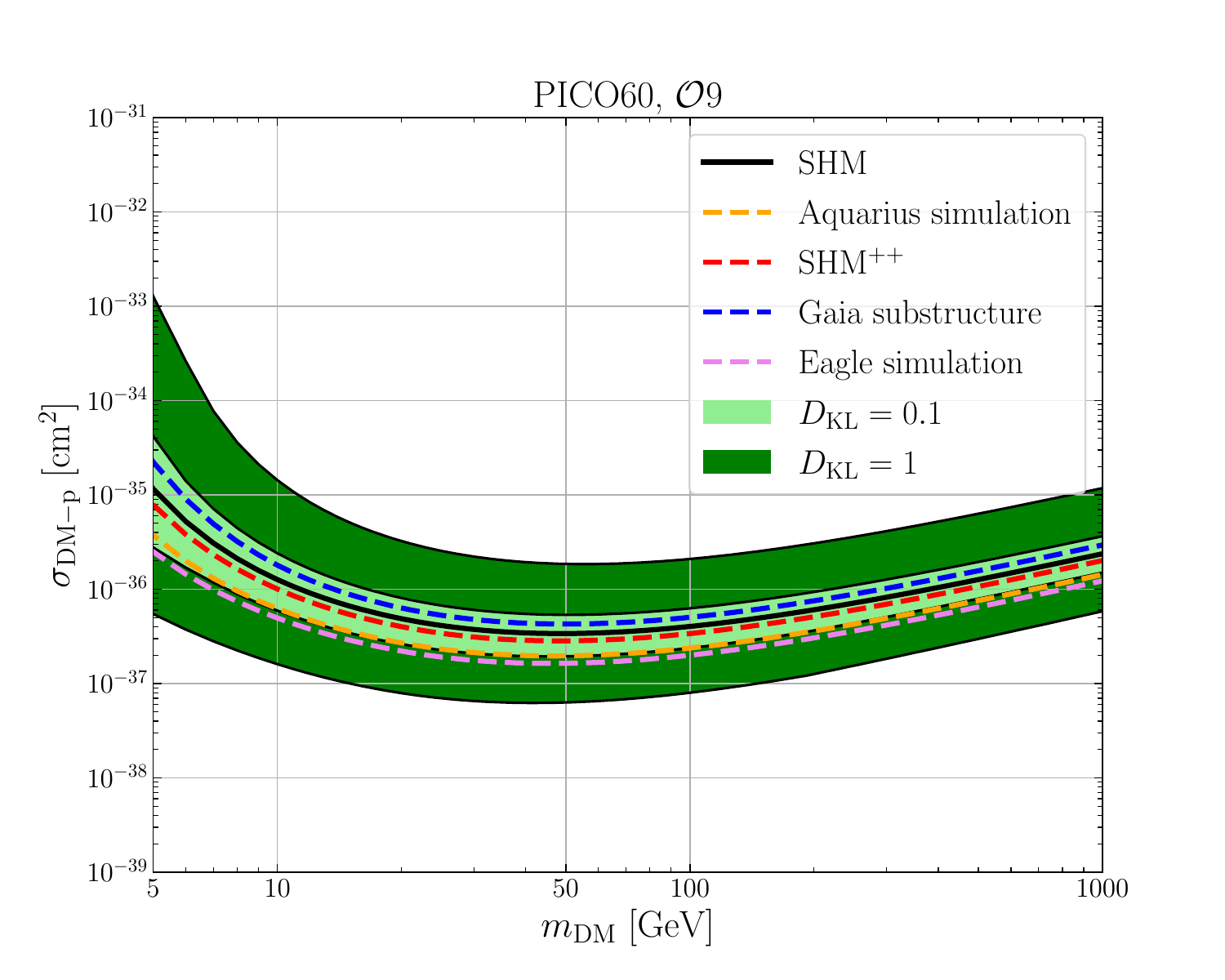}
\end{minipage}
\end{figure}

%%%%%%%%%%%%%%%%%%%%%%%%%%%%%%%%%%%%%%%%%%%%%%%%%%%%%%%%%%%%%%%%%%%%%%%%%%
%#######################################

\section{Conclusions}\label{sec:conclusions}

Direct detection experiments are entering an era where the coherent scattering of solar neutrinos becomes an irreducible background for a dark matter signal. In reality, this is only true for coherent, elastic, spin-independent interactions. Dark matter may interact through any of the operators of the non-relativistic effective theory, whose associated cross sections present different dependencies on the incoming dark matter velocity and momentum transfer of the scattering process. These velocity and momentum dependencies can amplify the role of astrophysical uncertainties in the local dark matter distribution and significantly affect the ensuing experimental limits.

\begin{table}[t!]
\centering

\label{tab:NR_bilinear_UV_astro}
\begin{tabular}{p{1cm} p{5cm} p{5.0cm} p{4.0cm}}
\hline\hline
NR
& Relativistic bilinear
& UV dark matter example
& Impact of astrophysics
\\
\hline
$\mathcal{O}_1$
& $\bar\chi\chi\,\bar q q$;
  $\bar\chi\gamma^\mu\chi\,\bar q\gamma_\mu q$
&  Higgs portal, $Z^{\prime}$, \textit{e.g} \cite{Goodman:1984dc}
& Low ($\lesssim$1 o.f.m)
\\
$\mathcal{O}_6$
& $\bar\chi i\gamma^5\chi\,\bar q i\gamma^5 q$
& ALP mediator, \textit{e.g} \cite{Freytsis:2010ne}
& Moderate ($\sim$1-2 o.f.m)
\\
$\mathcal{O}_5$
& $\bar\chi\sigma^{\mu\nu}\chi\,F_{\mu\nu}$
& Dipole, composite, \textit{e.g} \cite{Pospelov:2000bq}
& High ($\sim$2-3 o.f.m)
\\
$\mathcal{O}_8$
& $\bar\chi\gamma^\mu\gamma^5\chi\,\partial^\nu F_{\mu\nu}$
& Majorana (Anapole), \textit{e.g} \cite{Fitzpatrick:2010br}
& High ($\sim$2--3 o.f.m)
\\
$\mathcal{O}_{13}$
& $\left(\bar\chi \gamma^\mu\gamma^5 \chi\right)\,\partial_\mu\!\left(\bar q q\right)$
& Pseudo-Goldstone, \textit{e.g} \cite{Gross:2017dan}
& High ($\sim$3 o.f.m)
\\
$\mathcal{O}_{14}$
& $\left(\bar\chi \sigma^{\mu\nu}\chi\right)\,\partial_\mu\!\left(\bar q \gamma_\nu\gamma^5 q\right)$
& UV-suppressed EFT, \textit{e.g} \cite{Brod:2017bsw}
& Very high ($\sim$3--4 o.f.m)
\\
\hline\hline
\end{tabular}
\caption{Some examples of correspondence between non-relativistic operators, their typical relativistic bilinear origins, representative UV dark matter model classes, and the order-of-magnitude impact of astrophysical
uncertainties on inferred direct-detection limits near the kinematical threshold of XENONnT and PICO60. o.f.m denotes "order of magnitude".}
\end{table}

Here we have provided a unified and model-independent quantification of these uncertainties across the full non-relativistic effective field theory basis for the XENONnT and PICO60 experiments. By constraining the deviation of the true velocity distribution from the Maxwell-Boltzmann form through the Kullback-Leibler divergence, we formulated a convex optimization problem whose solution yields the most conservative and most aggressive upper limits on the Wilson coefficients compatible with a given information-theoretic bound \cite{Herrera:2024zrk}. This approach requires no parametric assumptions about the functional form of the velocity distribution and naturally accommodates substructure and unvirialized components.

We revealed a clear hierarchy in the sensitivity of different operators to astrophysical uncertainties. Operators such as $\mathcal{O}_1$ and $\mathcal{O}_4$, whose rates depend only on the first moment of the velocity distribution, are comparatively robust: for $D_{\rm KL}=0.1$, their limits vary only at the level of a factor of a few relative to the Standard Halo Model. In contrast, operators dominated by higher powers of the dark matter velocity or by momentum-suppressed nuclear responses, such as $\mathcal{O}_5$, $\mathcal{O}_7$, $\mathcal{O}_8$, $\mathcal{O}_{14}$, and $\mathcal{O}_{15}$, can exhibit uncertainties of up to two or three orders of magnitude near the experimental threshold, precisely where the background from solar neutrinos becomes sizable. The pattern observed across operators is well captured by a moment-based interpretation: operators probing raw-velocity higher moments of the velocity distribution (associated with the variance and skewness) inherit a stronger dependence on the high-velocity tail and are therefore less stable under deviations from the SHM.

Table~\ref{tab:NR_bilinear_UV_astro} illustrates how this hierarchy maps onto representative relativistic bilinear structures and UV dark matter scenarios, highlighting that interactions generated by derivative, dipole, or higher dimension tensor operators are intrinsically more sensitive to astrophysical uncertainties.

The information-theoretic framework developed here can be applied directly to reinterpret existing and upcoming experimental results, in particular in the regime where neutrino induced recoils and dark matter induced recoils become increasingly degenerate. The method is general and can be readily extended to multi-target analyses, inelastic scattering scenarios, the Migdal effect, or the non-relativistic effective field theory of dark matter-electron interactions. Incorporating nuclear-physics uncertainties and joint constraints from future experiments with different target compositions and solar neutrino measurements are important directions for future work. 

More generally, our technique could be applied to other physics problems where a given physical observable presents a functional form resembling the statistical moments of a given uncertain distribution.
By maintaining a compromise between analyticity and statistical power, this framework provides a complementary perspective to data driven and deep learning based methods, particularly in contexts where interpretability and physical transparency are essential.

\subsection*{Acknowledgments}
We are grateful to Gaurav Tomar, Stefano Scopel and Andreas Rappelt for useful discussions. Our work is supported by the Neutrino Theory Network Fellowship with contract number 726844, and by the U.S. Department of Energy under award number DE-SC0020262. This manuscript has been authored by FermiForward Discovery Group, LLC under Contract No. 89243024CSC000002 with the U.S. Department of Energy, Office of Science, Office of High Energy Physics.

\appendix
\appendix
\section{Relation between velocity-weighted integrals and statistical moments}
\label{app:moments}

In this appendix we clarify the relation between the velocity-weighted integrals that appear in the direct detection event rate and the standard statistical moments of a probability distribution. This discussion is intended to make precise the sense in which the velocity dependence of different non-relativistic operators can be associated with familiar statistical descriptors, while highlighting the limitations of this analogy in the presence of experimental thresholds.

Throughout the main text, the dependence of the event rate on the dark matter velocity distribution can be expressed schematically in terms of integrals of the form
\begin{equation}
I_n(v_{\min}) \equiv \int_{v_{\min}}^{\infty} dv \, v^n \, f(v),
\label{eq:In_def}
\end{equation}
where \(f(v)\) denotes the angular averaged dark matter speed distribution in the laboratory frame and \(v_{\min}\) is the minimum velocity required to produce a detectable nuclear recoil. The integer \(n\) depends on the velocity and momentum structure of the corresponding non-relativistic operator and the associated response functions.

The quantities \(I_n(v_{\min})\) are truncated raw moments of the speed distribution. In general, they are neither normalized nor defined over the full support of the distribution, and therefore do not correspond directly to statistical moments in the usual sense. To establish a precise connection with standard statistical descriptors, it is useful to define a normalized conditional speed distribution restricted to velocities above the experimental threshold,
\begin{equation}
p(v \mid v \geq v_{\min}) \equiv \frac{f(v)}{I_0(v_{\min})},
\qquad v \geq v_{\min},
\label{eq:conditional_dist}
\end{equation}
which satisfies
\begin{equation}
\int_{v_{\min}}^{\infty} dv \, p(v \mid v \geq v_{\min}) = 1.
\end{equation}

The raw moments of this conditional distribution are then given by
\begin{equation}
m_n(v_{\min}) \equiv \int_{v_{\min}}^{\infty} dv \, v^n \, p(v \mid v \geq v_{\min})
= \frac{I_n(v_{\min})}{I_0(v_{\min})}.
\label{eq:raw_moments_conditional}
\end{equation}

The usual central moments of the conditional distribution can now be expressed exactly in terms of the velocity-weighted integrals \(I_n(v_{\min})\).
The conditional mean speed is
\begin{equation}
\mu(v_{\min}) = m_1(v_{\min}) = \frac{I_1(v_{\min})}{I_0(v_{\min})}.
\label{eq:mean_conditional}
\end{equation}

The conditional variance is
\begin{equation}
\sigma^2(v_{\min}) = m_2(v_{\min}) - m_1^2(v_{\min})
= \frac{I_2(v_{\min})}{I_0(v_{\min})}
- \left( \frac{I_1(v_{\min})}{I_0(v_{\min})} \right)^2 .
\label{eq:variance_conditional}
\end{equation}

The third central moment of the conditional distribution is
\begin{align}
\mu_3(v_{\min})
&= \int_{v_{\min}}^{\infty} dv \, \bigl(v - \mu(v_{\min})\bigr)^3 \,
p(v \mid v \geq v_{\min}) \nonumber \\
&= m_3(v_{\min}) - 3 m_2(v_{\min}) m_1(v_{\min})
+ 2 m_1^3(v_{\min}) \nonumber \\
&= \frac{I_3(v_{\min})}{I_0(v_{\min})}
- 3 \frac{I_2(v_{\min}) I_1(v_{\min})}{I_0^2(v_{\min})}
+ 2 \left( \frac{I_1(v_{\min})}{I_0(v_{\min})} \right)^3 .
\label{eq:third_central_moment}
\end{align}

From this, the conditional skewness is defined as
\begin{equation}
\gamma_1(v_{\min}) =
\frac{\mu_3(v_{\min})}{\sigma^3(v_{\min})},
\label{eq:skewness_conditional}
\end{equation}

and analogous expressions can be derived for higher-order central moments. These equations demonstrate explicitly how the velocity-weighted integrals \(I_n(v_{\min})\) that appear in the scattering rate are related to the standard statistical descriptors of the normalized speed distribution above threshold.
Operators whose rates are dominated by different powers of \(v\) therefore exhibit indirect sensitivity to the conditional mean, variance, or skewness of the velocity distribution, through these algebraic relations.
It is clear that each operator probes a specific truncated raw moment \(I_n(v_{\min})\), with the corresponding sensitivity to familiar statistical descriptors emerging only after normalization and combination of multiple such integrals. This distinction becomes particularly important near threshold, where the truncation at \(v_{\min}\) enhances sensitivity to the high-velocity tail of the distribution.

\printbibliography

\end{document}